%% file: lxepur.tex
\RequirePackage{fix-cm}
\documentclass[twocolumn,final,epjc3]{svjour3}  

\smartqed  
\usepackage{hyperref}
\usepackage{graphicx}
\usepackage{multirow}
\usepackage{amsmath,amssymb,amsfonts}
\usepackage{mathrsfs}
\usepackage{xcolor}
\usepackage{textcomp}
\usepackage{manyfoot}
\usepackage{booktabs}
\usepackage{algorithm}
\usepackage{algorithmicx}
\usepackage{algpseudocode}
\usepackage{listings}
\usepackage{rotating}
\usepackage{gensymb}
\usepackage{xspace}             
\usepackage[separate-uncertainty=true]{siunitx}
\usepackage{subcaption}
\usepackage{setspace}
\usepackage{wasysym} 
\usepackage{url}
\usepackage{cite} 
\usepackage{orcidlink}          

\usepackage[mathlines]{lineno}

\newcommand{\mmxe}{\ensuremath{M_\mathrm{Xe}}\xspace}
\newcommand{\venturiflow}{\ensuremath{\dot{V}_\mathrm{pur}}\xspace}
\newcommand{\pumpflow}{\ensuremath{\dot{V}_\mathrm{P}}\xspace}
\newcommand{\bypassflow}{\ensuremath{\dot{V}_\mathrm{B}}\xspace}
\newcommand{\radflow}{\ensuremath{\dot{V}_\mathrm{rad}}\xspace}
\newcommand{\pumpdp}{\ensuremath{\Delta P_\mathrm{P}}\xspace}
\newcommand{\venturidp}{\ensuremath{\Delta P_\mathrm{ven}}\xspace}
\newcommand{\dqdl}{\ensuremath{\frac{dq}{dl}}\xspace}
\newcommand{\dqdlt}{\ensuremath{dq/dl}\xspace}

\newcommand{\pressureat}[2]{\ensuremath{P_{\mathrm{\scriptscriptstyle #1{\text -}#2}}}}
\newcommand{\deltapof}[2]{\ensuremath{\Delta P_{\mathrm{\scriptscriptstyle #1{\text -}#2}}}}
\newcommand{\cvof}[2]{\ensuremath{C_{\mathrm{\scriptscriptstyle V, #1{\text -}#2}}}}
\newcommand{\yof}[2]{\ensuremath{y_{\mathrm{\scriptscriptstyle #1{\text -}#2}}}}
\newcommand{\flowcoefficient}{\ensuremath{C_{\scriptscriptstyle V}}\xspace}
\newcommand{\otwo}{\ensuremath{\mathrm{O}_2}\xspace}
\newcommand{\water}{\ensuremath{\mathrm{H}_2\mathrm{O}}\xspace}
\newcommand{\lntwo}{\ensuremath{\mathrm{LN}_2}\xspace}
\newcommand{\rnbkg}{${}^{222}$Rn\xspace}

\newcommand{\kotwoval}{\SI[per-mode=power]{1.58e11}{\liter \per \mol \per \second}\xspace} 
\newcommand{\pmpressureval}{\SI{5.2}{\bar}\xspace} 
\newcommand{\pmtempval}{\SI{176}{\kelvin}\xspace} 
\newcommand{\lxedensityval}{\SI{2.869}{\kilo \gram \per \liter}\xspace} 
\newcommand{\negamtcaptured}{\SI{7.5}{\milli \gram}\xspace}  
\newcommand{\negefficiencyfit}{\SI{55}{\percent}\xspace}
\newcommand{\nTtotalLXeMass}{\SI{8.6}{t}\xspace}
\newcommand{\nTCryostatLXeMass}{\SI{8.5}{t}\xspace}

\DeclareSIUnit\inch{in}
\DeclareSIUnit\bar{bar}
\DeclareSIUnit\literperminute{LPM}
\DeclareSIUnit\standardliterperminute{SLPM}
\DeclareSIUnit \ppb{ppb}
\DeclareSIUnit \ppt{ppt}
\DeclareSIUnit \ppq{ppq}

\journalname{Eur. Phys. J. C}

\begin{document}

\title{The liquid-phase xenon purification system of the XENONnT dark matter experiment}

\input{Sep26-LXePur}

\date{Received: date / Accepted: date}

\onecolumn
\maketitle
\twocolumn

\begin{abstract}
The XENONnT dark matter experiment utilizes a dual-phase time projection chamber (TPC) housed in a cryostat filled with \nTCryostatLXeMass of ultra-pure liquid xenon. To achieve optimal sensitivity, free electrons generated by particle interactions are required to drift across the 1.5-meter detector under an electric field. Dissolved impurities, particularly oxygen (\otwo), can attenuate the resulting charge signal due to electron attachment. XENONnT demands an \otwo-equivalent concentration below 100 parts per trillion (ppt) to enable its dark matter sensitivity reach.

A novel liquid-phase xenon purification system was developed that continuously circulates liquid xenon through custom-designed, packed-bed sorbent purifier units at a volumetric flow rate of \num{\sim 2} liters per minute (\si{\literperminute}) at a cryogenic temperature around \SI{175}{\kelvin}. 
With a purification time constant of approximately \SI{25}{\hour}, the system reached the required purity within five days.
The purification strategy employs a two-step approach: For an initial cleanup, a purifier based on a copper catalyst on high-surface-area alumina support (Q5) removes impurities from the liquid xenon bulk with nearly \SI{100}{\percent} efficiency, ideal for initial detector filling or after long-term maintenance. However, the high \rnbkg emanation rate of the Q5 catalyst leads to increased background in the detector. In a second step, a purifier with sintered pellets of non-evaporable getter alloy (St707) ensures sustained purity with negligible background impact. The St707-based purifier exhibits a \negefficiencyfit efficiency, sufficient for normal operation. Following system commissioning, XENONnT achieved an \otwo-equivalent concentration in liquid xenon below \SI{19}{ppt} during its first science run. With this exceptional purity, over \SI{86}{\percent} of generated electrons survive a full drift across the TPC.
\end{abstract}
\keywords{Liquid xenon \and Liquid-phase purification \and Electronegative impurities \and Electron lifetime \and Non-evaporable getter \and Dark matter detectors}
\section{Introduction}
\label{sec:intro}
Dual-phase liquid xenon time projection chambers (TPCs) have revolutionized dark matter searches, enabling unparalleled sensitivity to Weakly Interacting Massive Particles (WIMPs)~\cite{xenon10:first,xenon100:combined,lux:complete,xenon1t:1ty,pandax2:full,PandaX4T:WIMP:2025,XENONnT:WIMP:2025,LZ:WIMP:2025}. Their success stems from the ease with which their liquid xenon (LXe) target mass can be scaled while maintaining exceptional low-energy (keV-scale) nuclear recoil detection efficiency. XENONnT, the largest and most recent TPC in the XENON Dark Matter project, is located at the INFN Laboratori Nazionali del Gran Sasso (LNGS) in Italy~\cite{xenonnt:instrument}.

Particle interactions within a TPC generate both primary scintillation photons and ionization electrons. A uniform electric field is applied to drift the ionization electrons upwards towards the liquid-gas interface. Upon reaching the surface, these electrons are extracted into the gaseous xenon (GXe) phase by a stronger electric field, where they are accelerated to produce secondary scintillation photons through electroluminescence~\cite{dolgoshein}. However, impurities dissolved in the LXe can hinder this process by capturing drifting electrons or absorbing both primary and secondary scintillation photons before they are detected by photomultiplier tubes (PMTs) arranged in arrays at the top and bottom of the TPC. These detrimental effects lead to attenuated signals, worsening the low-energy threshold of TPCs and increasing response variations within the sensitive LXe volume, ultimately impacting their sensitivity to potential WIMP interactions.

Water (\water) and molecular oxygen (\otwo) are two primary culprits behind the attenuation of scintillation and ionization signals in TPCs. \water's substantial light absorption cross-section at the scintillation wavelength of LXe (\SI{175}{nm})~\cite{mpimainz:uvdb,meg:water} and the inherent difficulty of removing it from vacuum systems, where a bakeout at high temperature is not possible~\cite{ohanlon}, contribute significantly to signal attenuation. On the other hand, \otwo's high electron attachment rate constant in LXe~\cite{bakale} leads to the capture of free electrons during their drift, further diminishing the signal. Both \water and \otwo contamination often originate from detector components that have been exposed to air eventually in contact with LXe or GXe. While extensive evacuation of the cryostat inner vessel enclosing the TPC using turbomolecular pumps prior to cooling and filling it with LXe helps to considerably reduce initial contamination, residual \water on metal surfaces and \otwo within porous materials remain unavoidable. These impurities continuously desorb from detector materials, accumulating in the LXe and GXe volumes of the cryostat. To maintain low impurity concentrations, continuous purification of these volumes is essential. Considering the comparative electron attachment rate constant of \otwo and the light absorption cross-section of \water, coupled with their typical desorption rates at LXe temperatures, achieving suitable \otwo concentrations poses a larger challenge.

When a significant number of electrons are lost to attachment, the recorded signals are no longer directly proportional to the energy deposited. This necessitates corrections based on the interaction depth within the TPC. The LXe purity is typically quantified by the ``electron lifetime'' $\tau_e$, defined as the characteristic time for the number of drifting electrons $N_e$ to be reduced by a factor $1/e$ due to attachment:
\begin{linenomath*}
\begin{equation} \label{eq:elifetime}
N_e(t) = N_e(0)\ e^{-t/\tau_e}.
\end{equation}
\end{linenomath*}
The electron lifetime is directly linked to the concentration $C_i$ and attachment rate constants $k_i$ of impurities through the following equation:
\begin{linenomath*}
\begin{equation} \label{eq:rateconstant}
\tau_e = \frac{1}{\sum k_i C_i} \approx \frac{1}{k_{\otwo} C_{\otwo}},
\end{equation}
\end{linenomath*}
where $C_{\otwo}$, the \otwo-equivalent concentration, serves as a standard measure of purity due to \otwo's dominant role in attenuating charge signals in TPCs~\cite{bakale}. Throughout this work, the \otwo-equivalent mole fraction $x_{\otwo}$, often expressed in parts per trillion (ppt), is used interchangeably with electron lifetime based on the relationship:
\begin{linenomath*}
\begin{equation}\label{eq:elconversion}
    \frac{1}{\tau_e} = k_{\otwo} C_{\otwo} = k_{\otwo} x_{\otwo} \frac{\rho} {\mmxe} = \frac{x_{\otwo}}{\SI{289}{\milli \second \cdot ppt}}.
\end{equation}
\end{linenomath*}
Here, the LXe density $\rho=\lxedensityval$ is calculated for the approximate operating pressure (\pmpressureval) and temperature (\pmtempval) of the purity monitor (described in Sec.~\ref{subsec:pm}). The molar mass of xenon \mmxe is \SI{131}{\gram \per \mole}. The rate constant $k_{\otwo}=\kotwoval$ is adopted from Ref.~\cite{bakale} for the purity monitor's \SI{100}{\volt\per\centi\meter} drift field utilized throughout this study.

In larger TPCs, where drift times are longer, maintaining acceptable ionization signal losses requires higher electron lifetimes (lower \otwo concentrations). While intuitively, the decreased surface-to-volume ratio of larger TPCs might suggest a potential advantage in achieving lower \otwo concentrations, this is usually not the case due to several factors. Polytetrafluoroethylene (PTFE), a commonly used material for efficient scintillation light reflection, frequently constitutes the primary source of \otwo desorption. As a structural component of the TPC, PTFE's mass must be increased to support (counteract) the heavier bottom PMT array (buoyancy), leading to a larger surface area from which \otwo can desorb. Moreover, thicker structural components hinder the outdiffusion of \otwo molecules, extending the required evacuation time. Abbreviating this evacuation process can inadvertently increase the desorption rate per unit surface area. Finally, stringent cleanliness protocols become more challenging to implement with larger-scale components, making it harder to achieve low \otwo concentrations.

Traditionally, TPCs employ heated getter gas purifiers to achieve low impurity concentrations through continuous recirculation of their target~\cite{xenon100:instrument,lux:instrument,pandax:cryo,pandax4t:cryo,xenon1t:instrument,lz:instrument}. These purifiers utilize a tri-metal alloy (zirconium, vanadium, and iron) to permanently capture common impurities, including \water, \otwo, $\mathrm{N}_2$, $\mathrm{CO}$, $\mathrm{CO}_2$, and $\mathrm{CH}_4$. To accommodate mass flow rates up to \num{100} standard liters per minute (\si{\standardliterperminute}\footnote{defined at standard conditions for temperature (\SI{273.15}{K}) and pressure (\SI{1}{atm)}.}), two-phase heat exchangers are integrated to vaporize (recondense) the outgoing LXe (incoming GXe) flow between the TPC and the purifier, thereby minimizing the external cooling power needed to purify the LXe target in the gaseous phase. XENON1T~\cite{xenon1t:instrument}, the predecessor to XENONnT, successfully achieved an exchange of its \SI{3.2}{\tonne} volume every 7 days using this technique, resulting in an electron lifetime of approximately \SI{700}{\micro \second} during its primary science run~\cite{xenon1t:1ty}. Following a pump upgrade, the electron lifetime was enhanced to \SI{1}{\milli \second} with a 1.4-fold increase in flow rate. A larger LXe target mass, coupled with the required ultra-low \otwo concentration, necessitates an increased mass flow capacity for the gas purifier. This translates into reduced flow resistance in the process piping and valves, more efficient heat exchangers, and gas pumps capable of delivering higher pressure differentials at the elevated flow rates.

A target electron lifetime of \SI{1}{\milli \second} was derived from XENONnT's projected WIMP sensitivity~\cite{xenonnt:mc} together with the dedicated purity study of Ref.~\cite{nobuthesis}. Incorporating a conservative safety factor of 3, the design electron lifetime was set at approximately \SI{3}{\milli \second}, requiring \otwo-equivalent concentrations below \SI{100}{ppt}. To achieve this, XENONnT stands out as the first dark matter TPC-based experiment to adopt a novel cryogenic liquid-phase purification (LXePUR) approach for its \nTtotalLXeMass LXe target. While liquid-phase purification has demonstrated its effectiveness in removing \water and \otwo from large liquid argon TPCs~\cite{icarus:purity2,icarus:liqpur,dune:cdrvol4,argoneut:liqpur} and \water from LXe in the MEG~\cite{meg:liqpur} and XMASS~\cite{ABE201378} experiments, this marks the first application of this method for removing \otwo in a TPC-based dark matter direct detection experiment. Cryogenic liquid purification systems developed prior to the one for XENONnT, often employing molecular sieves and highly dispersed copper catalysts, typically require substantial quantities of sorbents/reactants, and suffer from significant \rnbkg emanation rates. These limitations render them unsuitable for use in purifiers for TPC-based low-background experiments, particularly those focused on dark matter direct detection, where \rnbkg decay products constitute the primary electronic recoil background~\cite{xenonnt:sr0wimp}.

The XENONnT TPC is housed within a vacuum-insulated double-walled stainless steel cryostat, which is suspended at the center of the water tank, a \SI{\sim 10}{\meter} diameter and \SI{\sim 10}{\meter} high structure, serving as a passive shield and an active muon and neutron veto system~\cite{xenonnt:instrument}. XENONnT reuses the XENON1T infrastructure, including the cryogenic, gaseous purification (GXePUR), krypton distillation, and storage systems, located in the adjacent three-story service building. In addition to the LXePUR system, two new systems have been integrated: a high-flow radon removal system based on cryogenic distillation~\cite{murra:2022} and a solid/gas xenon storage system with a \SI{10}{t} capacity. The XENONnT cryostat's inner vessel and vacuum insulation are connected to the cryogenic system, situated on the third floor of the service building, via the original XENON1T double-walled stainless steel cryogenic pipe, referred to as the `cryopipe'. The GXePUR system, upgraded with two redundant magnetically-coupled piston pumps~\cite{brown:2018}, continues to purify the cryostat's LXe volume through two-phase heat exchangers. This purification process is often termed `evaporated-LXe' purification to distinguish it from the LXePUR system.

The Xenon Cryogenic Liquid Purification Setup (Xeclipse)~\cite{xeclipse}, a dedicated setup located at Columbia University and designed for liquid-phase xenon purification studies, was instrumental in identifying effective sorbent materials. Xeclipse test results, combined with precise measurements of the \rnbkg emanation rate from these sorbents~\cite{xenonnt:radiopurity}, provided the foundation for designing the LXePUR system's purifiers. 

In this paper, the impurity removal efficiency $\epsilon$  of a purifier with length $L$ is defined by the following equation:
\begin{linenomath*}
\begin{equation}
	\epsilon \equiv 1-\frac{x_{\mathrm{pur}} (z=L)}{x_{\mathrm{pur}} (z=0)},
	\label{eq:purifier_eff}
\end{equation}
\end{linenomath*}
where $x_{\mathrm{pur}} (z)$ represents the mole fraction of the impurity in the LXe (or GXe) stream after traversing a distance $z$ within the purifier. In general both $\epsilon$ and $x_{\mathrm{pur}}$ are also time-dependent.

Following the approach outlined in~\cite{xeclipse}, a simplified model is considered describing the removal of impurities from a single, continuously purified LXe volume. The rate of change in the impurity mole fraction $x$ within a volume containing a total quantity of LXe $n$ can be expressed as:
\begin{linenomath*}
\begin{equation} \label{eq:purity_model}
	n\frac{dx}{dt} = \Lambda -\frac{\epsilon n}{\tau} x,
\end{equation}
\end{linenomath*}
where $\Lambda$ represents the molar flow rate of impurities entering the LXe volume that is being processed in a time $\tau$ by a purifier with impurity removal efficiency $\epsilon$. At early times, starting from a high initial value, the molar fraction improves exponentially with purification time constant $\tau_{\mathrm{eff}} \equiv \tau/\epsilon$ and, at late times, reaches an equilibrium
\begin{linenomath*}
\begin{equation} \label{eq:equilibium_purity}
	x_{\infty} = \frac{\Lambda \tau}{n \epsilon} = \frac{\Lambda \tau_{\mathrm{eff}}}{n}.
\end{equation}
\end{linenomath*}
The purification time constant $\tau_{\mathrm{eff}}$ and the equilibrium mole fraction of impurities $x_{\infty}$ are two useful basic metrics to discuss the performance of a given purification system.

This paper provides a comprehensive overview of the XENONnT cryogenic LXe purification system, detailing its design (Sec.~\ref{sec:design}) and measured performance in terms of achievable flow rates, thermal efficiency, and its ability to purify the experiment's \nTtotalLXeMass of LXe (Sec.~\ref{sec:performance}).
\section{System design}
\label{sec:design}
The LXePUR for XENONnT was designed so that the experiment could achieve and maintain equilibrium \otwo concentrations in LXe at least lower than 100 ppt ($\simeq$ 3 ms electron lifetime) for extended periods of time. A number of features were therefore incorporated into the design to achieve this goal.

First and foremost was the capability of the system to remove impurities released in the LXe target with a short enough time constant and with a high enough efficiency to achieve the concentration goal. Based on the experience acquired in the operation of XENON100 and XENON1T, the design was predicated on the assumption that the crucial impurity to remove was \otwo. Furthermore, the rate of release of \otwo into the LXe target was assumed to be dominated by the outdiffusion from plastic materials that make up the TPC, mainly PTFE. A minimum LXe purifier \otwo removal efficiency requirement of $\sim$50\% and a maximum LXe purification volumetric flow requirement of \num{\sim 5} liters per minute (\si{\literperminute}\footnote{Note that throughout this work, liquid xenon volumetric flows are expressed in liters per minute (LPM), while gaseous xenon mass flows are expressed in standard liters per minute (SLPM), unless otherwise specified.}) were obtained by scaling the rate of release of \otwo measured in the early operation phase of XENON1T (\SI[per-mode=symbol]{\sim 0.1}{\milli\gram\per\day}) by the relative increase in total LXe mass in XENONnT (factor of 2.7). The use of PTFE as a structural element of the XENON TPCs was the main reason that motivated scaling with the volume instead of the surface-to-volume ratio.

The second feature that had a major impact on the design was the capability of the system to support continuous operation. At the most basic level, this is enabled by the use of two redundant cryogenic liquid pumps and two redundant LXe purifiers. A number of additional details are important, given that the system is purifying an ultra-clean cryogenic liquid. Cryogenic liquid pumps require periodic maintenance, such as rotor bearing replacement, and to do so while the system is operating with the second pump necessitates that the heat influx into the system from the pump being maintained be minimized. Even more important is eliminating the potential for the introduction of impurities when the original pump is brought back into service. Similar constraints dictate some design choices for the redundant LXe purifiers. The need to regenerate or replace a purifier while the system is in operation also implies minimizing the heat influx or the possibility of impurities being introduced into the system from these operations.

A third feature, the capability of the system to measure the concentration of impurities at a precise level, rapidly, and independently of the TPC also had a significant influence on the design. The solution adopted was to include a purity monitor in the LXe flow path. This provided the added benefit that it could potentially be used as a diagnostic tool for the purifiers.

The following subsections describe the various components of the XENONnT LXePUR system providing detailed descriptions and, where applicable, explaining the rationale behind the design choices and their connection to the basic features discussed above.
\subsection{Purification loop}
\label{subsec:loop}
\begin{figure*}[!h]
	\centering
	\includegraphics[width=\linewidth]{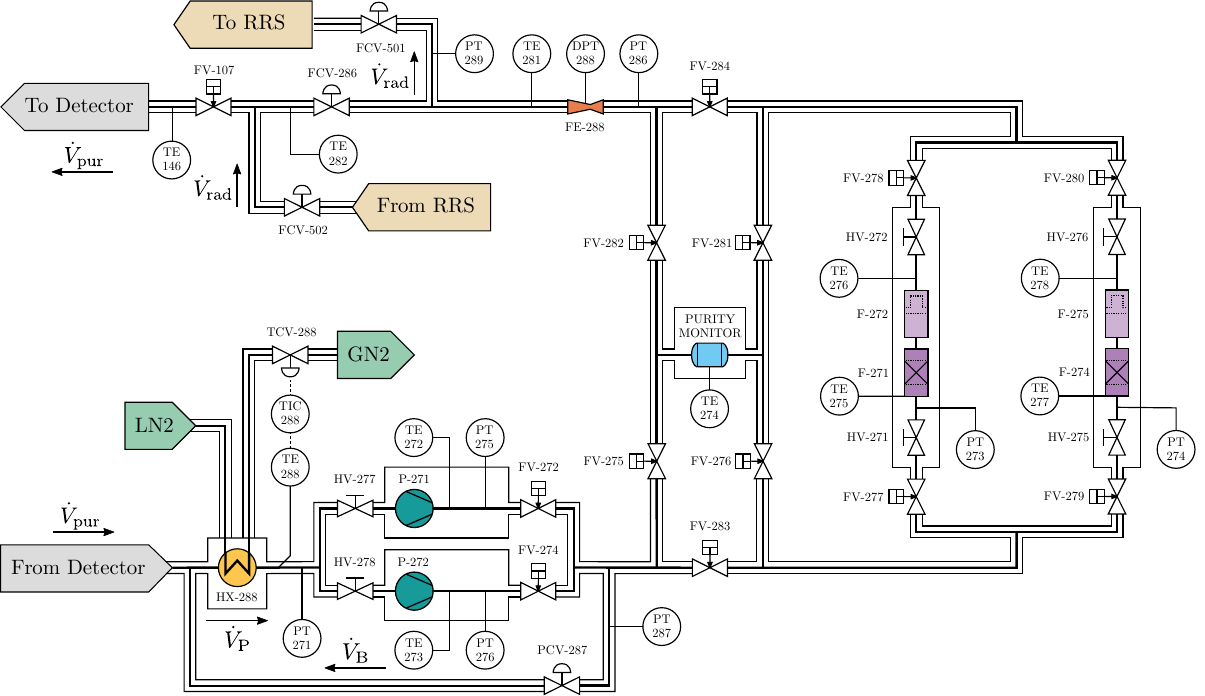}
 	\caption{Simplified process and instrumentation diagram of the XENONnT LXePUR system. The two redundant LXe pumps (P-271/272, teal) drive the LXe flow through the purification loop (\venturiflow). The $\mathrm{LN_2}$-LXe heat exchanger (HX-288, yellow) subcools the LXe flow upstream of the pumps and allows the system to operate with net negative heat input to the cryostat. The LXe flow through the operating pump (\pumpflow) and the bypass flow (\bypassflow) are larger than the purification loop flow. The two redundant purifiers (F-271/274, dark purple) in combination with particulate filters (F-272/275, light purple) capture impurities and particulates in the LXe stream. The purity monitor (blue) measures the electronegative impurity concentration, either from the detector or after the purifier units. The Venturi tube (FE-288, orange) measures the volumetric flow of the purification loop. A fraction of the purification loop flow can be diverted to the radon removal system (RRS, \radflow).}
 	\label{fig:pid}
\end{figure*}

\begin{figure*}[!h]
    \centering
    \includegraphics[width=\linewidth]{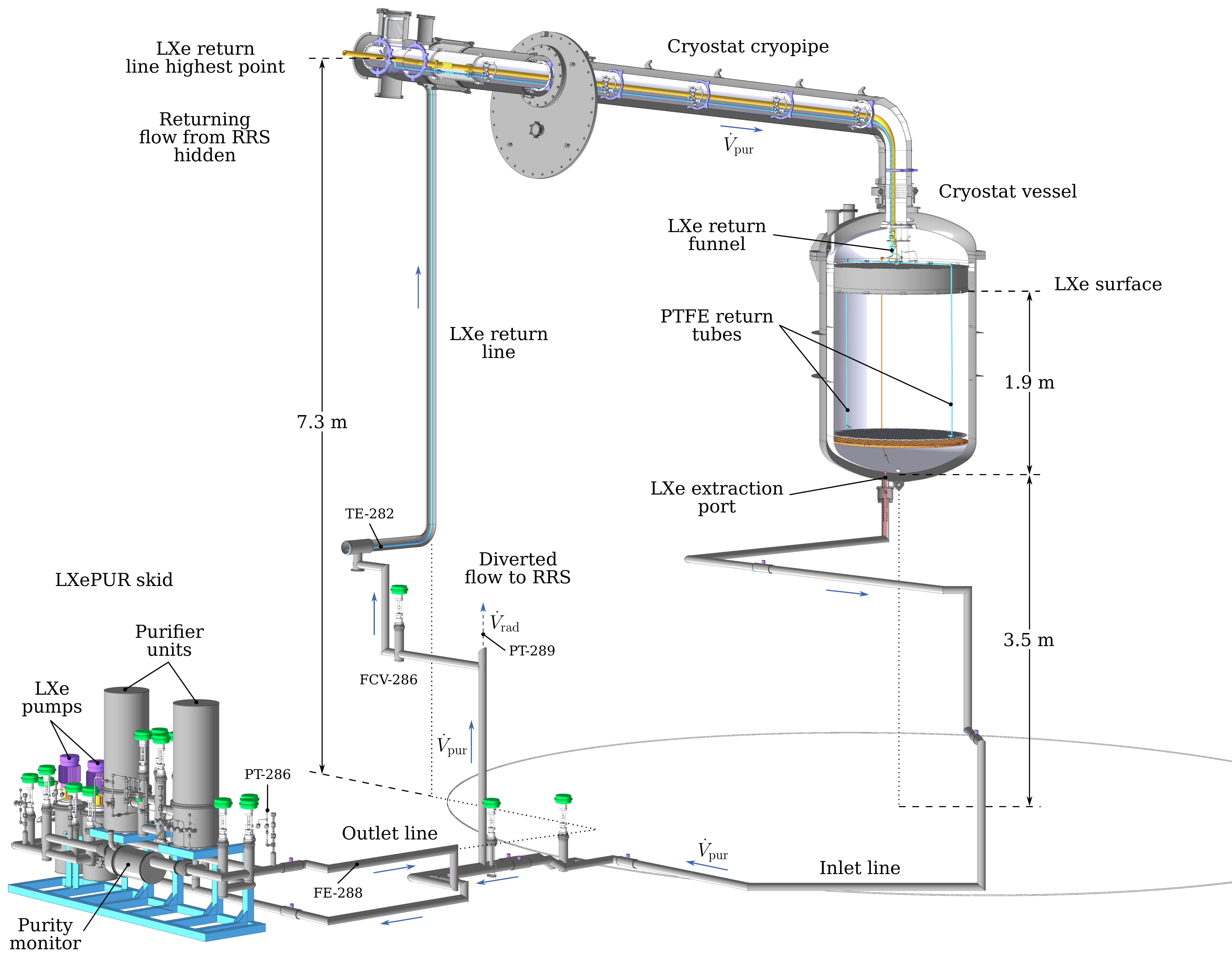}
    \caption{3D CAD rendering of the complete purification loop of the XENONnT LXePUR system. LXe is extracted from a port (red tube) at the bottom of the cryostat vessel, flows through and out of the water shield (circle footprint) via the inlet line to the LXePUR skid (lower left). Cryogenic liquid pumps (purple, only electrical motor visible) drive the LXe flow through the purity monitor, and the purifier units (gray vertical cylinders, only vacuum enclosure visible) then back to the detector via the LXePUR system outlet line and the LXe return line (turquoise tube), part of which runs in the cryostat cryopipe. The LXe return line is directed to a funnel immediately above the TPC, the outlet of which is directly connected to the TPC volume (enclosing PTFE panels not shown for clarity) via two diametrically opposed PTFE tubes (turquoise tubes). A number of cryogenic valves (green, pneumatic actuators visible) control various aspects of the operation of the LXePUR system. The tube-in-tube HE (yellow, LXe return to TPC; orange, LXe TPC outlet) that evaporates (condenses) the outgoing LXe (returning GXe) to (from) the GXePUR system (Sec.~\ref{sec:intro}) is also shown and shares the same return PTFE tubes to the TPC.}
    \label{fig:loop}
\end{figure*}

\begin{figure*}[!h]
    \centering
    \includegraphics[width=\linewidth]{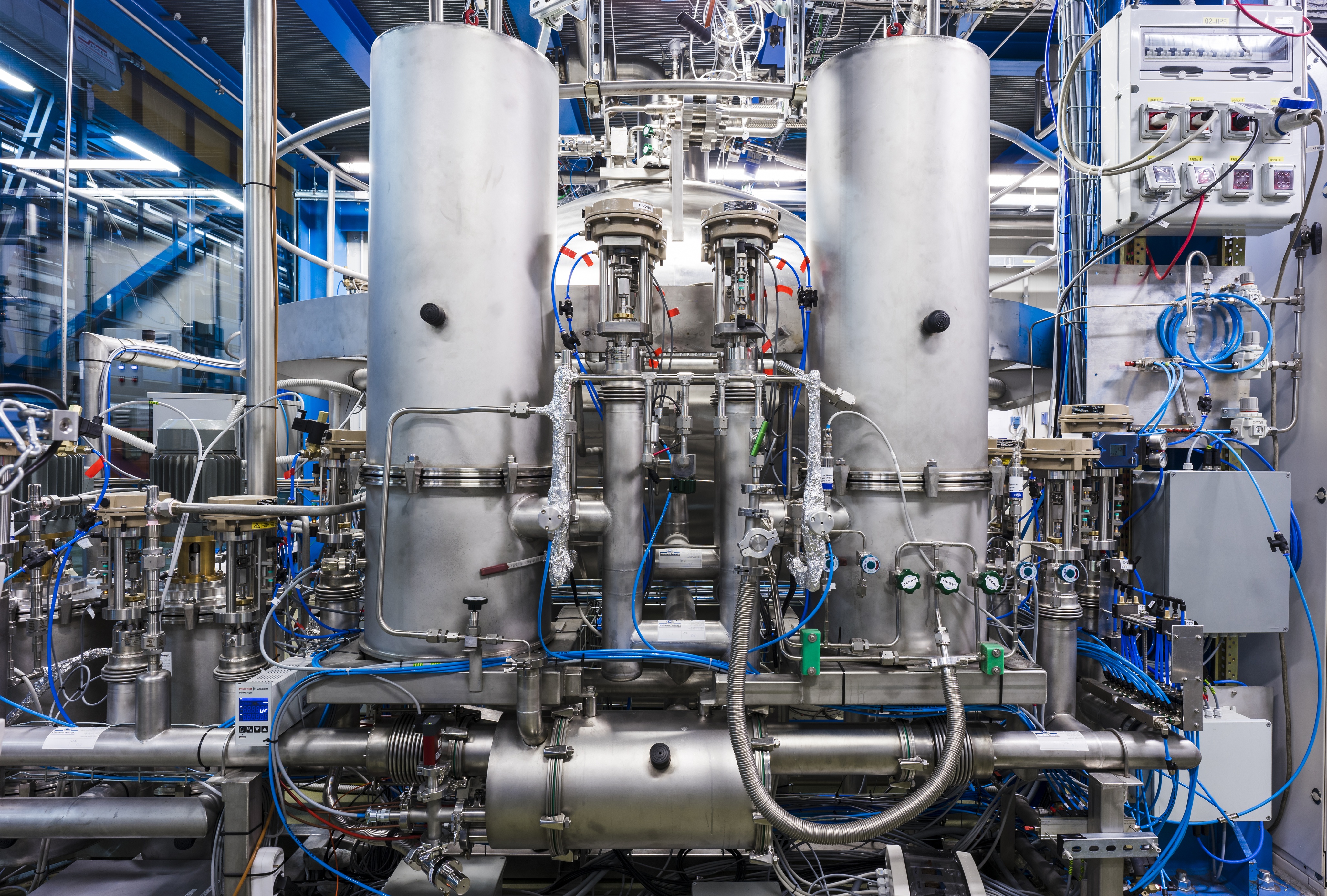}
    \caption{Picture of the XENONnT LXePUR system skid on the ground floor of the service building in front of the recovery and storage system for XENON (ReStoX). The two large vertical SS cylindrical vessels house the two redundant LXe purifier units and are equipped with flanges to access or replace the purifiers. The horizontal cylindrical vessel (bottom center) contains the purity monitor and also provides access via flanges and sealed sliding covers with bellows. The pump motors (gray) and their associated vacuum vessels are visible on the left behind vacuum-jacketed piping and valves.}
    \label{fig:skid}
\end{figure*}

The XENONnT LXePUR system is shown schematically in Fig.~\ref{fig:pid} and the physical layout of the complete purification loop is depicted in Fig.~\ref{fig:loop}. The LXePUR system is fed LXe from a port at the bottom of the XENONnT cryostat vessel and via a 21~m-long inlet line that runs through the water tank surrounding the cryostat, and into the ground floor of the XENONnT service building where the system is located~\cite{xenonnt:instrument}. The height difference between the cryostat port and the LXePUR system inlet is approximately \SI{3.5}{\meter}. Upon entering the LXePUR system, the LXe is subcooled by a liquid nitrogen (\lntwo) heat exchanger (HX-288) and then driven by one of two cryogenic liquid pumps (P-271/272) into one of two purifier units. Each purifier unit is composed of a LXe purifier (F-271/274) and a particulate filter (F-272/275). The purified LXe then passes through a Venturi tube (FE-288) to measure its volumetric flow (\venturiflow) before returning to the cryostat and into the TPC. As is common practice, the flow is calculated from a differential pressure measurement (DPT-288) between the entrance and the throat of the Venturi tube. The Venturi tube design follows ISO 5167-4 and was calibrated with water. The LXe out of the pump, or out of the purifier unit, can be directed through a purity monitor to measure its \otwo-equivalent impurities concentration. The two cryogenic liquid pumps, the purity monitor, and the two purifier units are all mounted on a skid. A picture of the LXePUR system skid is shown in Fig.~\ref{fig:skid}.

Part of the XENON1T LXe recovery line~\cite{xenon1t:instrument}, which ran from the bottom of the cryostat vessel, along the cryogenic double-wall pipe, and down to the ReStoX storage and recovery system, was re-purposed as the LXePUR system LXe return line from the ground floor of the XENONnT service building to the detector. The LXePUR system outlet line and part of the LXe return line, from the purifier units outlet to the cryostat double-wall pipe connection, which is the highest point, runs for \SI{15.5}{\meter}, with a height difference of \SI{7.3}{\meter}. The part of the LXe return line that runs in the cryostat double-wall pipe has a length of \SI{\sim 9}{\meter}. The LXe return line is directed to a funnel immediately above the TPC, the outlet of which is directly connected to the TPC volume via two diametrically opposed PTFE tubes. The funnel is opened to the cryostat GXe volume to act as a phase separator and prevent GXe from being pushed into the TPC. The two PTFE LXe return outlets are located halfway vertically between the bottom PMT array windows and the cathode electrode \cite{xenonnt:instrument}.


\begin{figure}[!h]
    \centering
    \includegraphics[width=\linewidth]{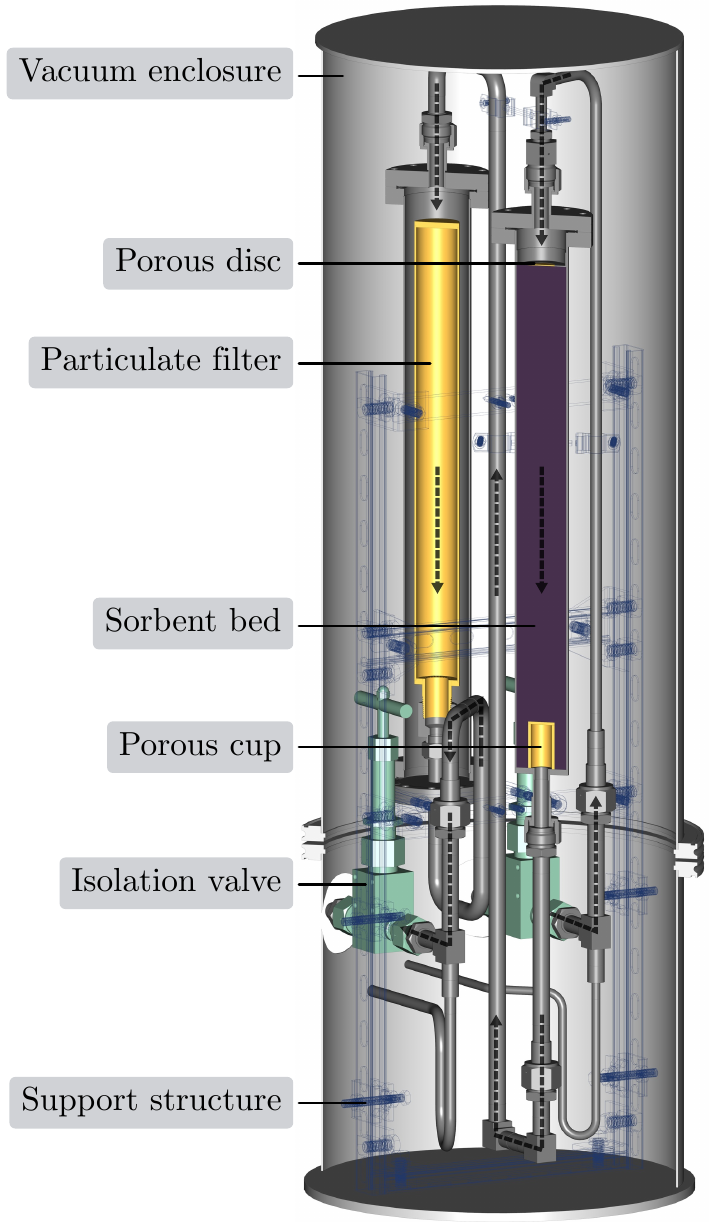}
    \caption{3D CAD rendering of the internal details of a LXe purifier unit: The LXe to be purified enters the vacuum enclosure of the purifier unit (gray) via a vacuum-jacketed tube and through an inlet secondary isolation valve (green, HV-271/275) flows upwards and then down through the sorbent bed (violet) of the purifier vessel (gray, F-271/274), passing through an inlet and outlet sintered SS 316L porous disc and cup (yellow), respectively. The purified LXe continues up, then down and through (from outside to inside) the particulate filter (yellow, F-272/275) before exiting the purifier unit through its outlet secondary isolation valve (green, HV-272/276). The cold purifier and particulate filter vessels and the tubing are rigidly held by an SS support frame (blue wireframe) with point thermal contacts to the room temperature enclosure to prevent vibrations of the assembly.}
    \label{fig:inside_purifier}
\end{figure}

The entire LXe flow path of the LXe purification loop piping is vacuum insulated, with the LXe-carrying inner tube covered with multilayer insulation (MLI) and supported by G10 (a fiberglass-epoxy composite material) spacers, as is common practice in the fabrication of vacuum-jacketed piping. However, the part of the LXe return line that runs in the cryostat cryopipe, itself vacuum insulated, is in direct contact with the GXe volume of the cryostat. For the vast majority of the purification loop, the piping used for LXe flow is \SI{0.75}{\inch} outer diameter (OD), \SI{0.065}{\inch} wall, electropolished stainless steel (SS) 316L tubing. The only exception is the upward flowing section upstream of each LXe purifier (see Fig.~\ref{fig:inside_purifier}) inside their respective vacuum enclosure, which is made of tubing with a \SI{0.5}{\inch} OD and a \SI{0.049}{\inch} wall, and the LXe return line to the detector that runs inside the cryostat cryopipe, which has an \SI{18}{\milli\meter} OD and a \SI{1}{\milli\meter} wall. All inner tube separable connections are made with \SI{0.75}{\inch} Swagelok VCR\textsuperscript{\textregistered}\ fittings.

A number of valves are used to select the desired pump, LXe purifier unit, and also to control various aspects of the operation of the system. The on/off valves (FV-2XX) used are pneumatically-actuated SS cryogenic bellows-sealed globe valves model 460PB from Thermomess~\cite{thermomess} with a flow coefficient (\flowcoefficient) of 4.2, chosen so that the total pressure drop at the maximum LXe flow would be within the capabilities of the pump. Two proportional valves (FCV-286, PCV-287), of the same basic model but equipped with an electropneumatic positioner, one of which (PCV-287) has a lower \flowcoefficient of 2.35, are used to control the overall flow of the LXe purification loop and the bypass flow of the cryogenic liquid pump (see section~\ref{subsec:pump}), respectively. All valves are vacuum insulated with a low heat-influx design and have an electropolished inner surface in contact with the LXe.

Several pressure transducers (PT-2XX) and temperature sensors (TE-2XX) are installed at various points along the LXe flow path to verify the proper operation of other components of the system, such as the LXe pump, as well as to monitor the thermal and flow characteristics of the system. The pressure transducers (Setra~\cite{setra} model 225 ultra-high purity metal sealed transducers) are coupled to the LXe flowing via room-temperature sensing tubes filled with GXe. The temperature sensors used are standard Pt100 temperature probes in a SS sheath welded onto a worm clamp and mounted on the outer surface of the LXe-carrying inner tube so as to ensure a low thermal contact resistance.

The XENONnT LXePUR system enables the operation of a high-flow radon removal system (RRS), based on cryogenic distillation~\cite{murra:2022}. This RRS effectively mitigates the impact of the most significant \rnbkg sources, those that directly emanate into the LXe volume of the detector. A desired fraction, typically \SI{0.4}{\literperminute}, of the LXe purification flow can be diverted to the RRS by the use of an arrangement of three proportional valves (FCV-501, FCV-286, and FCV-502). The triplet of proportional valves is also required to maintain the LXe volume contained in the RRS constant.

The instrumentation (pressure transducers, temperature sensors, etc) and the process equipment (cryogenic valves, LXe pumps, etc) of the LXePUR system are monitored and controlled, respectively, by the XENONnT slow control  system~\cite{xenonnt:instrument}. Multiple conditions (alarms) are defined to detect deviations from normal operation and potentially trigger automatic actions by the slow control system. For example, the differential pressure of the LXe pump in operation is continuously monitored and the pump is turned off if the differential pressure decreases below a low limit (\SI{0.5}{\bar}). Additional alarms monitor the degree to which the LXe at the outlet of the pump is subcooled and when the operation is too close to the pump's maximum allowable pressure. The slow control system automates both delicate and time-sensitive operations, such as the cooldown and initiation of the LXe pump, as well as routine tasks like high-voltage ramp up and measurements of the purity monitor.
\subsection{Cryogenic liquid pump}
\label{subsec:pump}
\begin{figure}[!h]
    \centering
    \includegraphics[width=\linewidth]{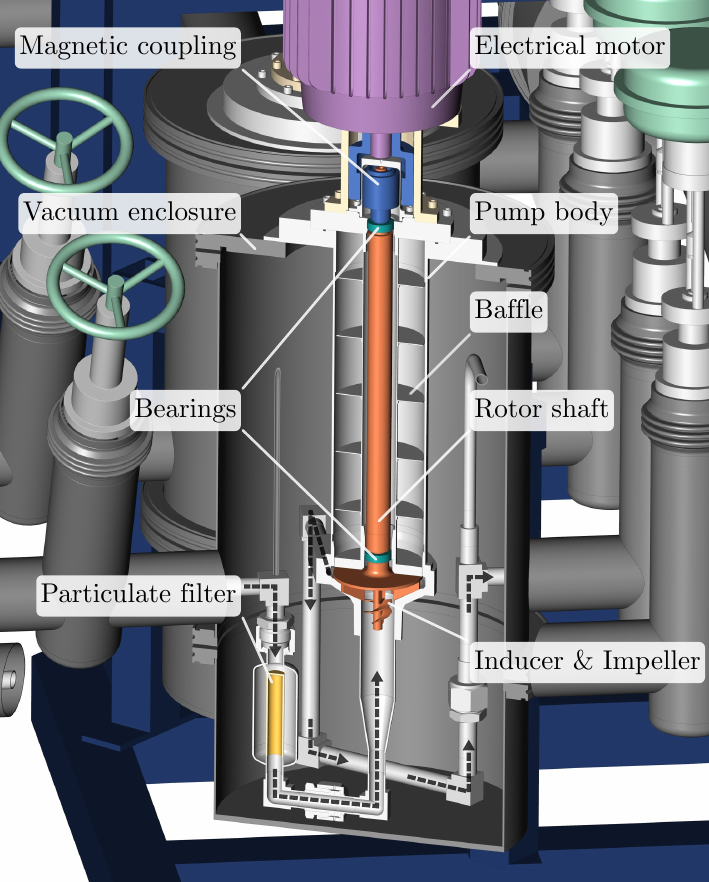}    
    \caption{3D CAD rendering of the internal details of a LXe pump: The LXe pump body (silver, external) is welded on the top flange of the pump vacuum enclosure (gray). The pump motor (violet) is magnetically-coupled (blue) to the rotor shaft (orange) to prevent impurities from the motor from contaminating the fluid being pumped. Two unlubricated bearings (teal) are seated into the fixed assembly (silver, internal) and allow the shaft with the inducer and impeller (orange, bottom) to rotate. Baffles (silver, internal) are used to reduce convective heat transfer in the GXe volume of the pump body along the shaft. Each pump is equipped with an inlet particulate filter (yellow).}
 \label{fig:inside_pump}
\end{figure}

The LXe purification loop flow is driven by two redundant Barber-Nichols~\cite{barbernichols} BNCP-32C-000 (partial emission) centrifugal cryogenic liquid pumps (P-271/272), with only one in operation at a time. This model features a magnetically-coupled shaft that completely isolates the motor and its windings from the rotating shaft and the impeller (see Fig.~\ref{fig:inside_pump}), and therefore from the fluid being pumped, minimizing the impurities introduced into the system upstream of the LXe purifiers. Two unlubricated bearings, one at the warm end, on top, one at the cold end, and on the bottom, allow the shaft to rotate, along with the impeller which is fixed at the cold end. Lubricants are avoided to minimize the impurity burden. The rotating shaft is made of a thin-walled tube to minimize the heat influx into the cryogenic liquid being pumped. The fixed part of the assembly, on which the bearings are resting, can be lowered into the pump body. It closes the volume containing the GXe, and the LXe being pumped, with a room temperature flange sealed with a Viton o-ring. It is also equipped with baffles to reduce convective heat transfer in the GXe volume of the pump body. The pump motor is mounted on top and is controlled by a Toshiba VF-S15 variable frequency drive.

Each pump body is installed in an independent SS vacuum-insulated vessel and covered with MLI.  Due to space constraints, the vacuum insulation volume is shared between the two pumps. Each pump is isolated by a manual inlet valve (HV-277/278 for P-271/272, respectively) at its inlet and an on/off pneumatic valve at its outlet (FV-272/274, respectively). It is, therefore, possible to dismount and maintain one pump while the other one is in operation. Tubing connections are foreseen (not shown in Fig.~\ref{fig:pid} for simplicity but partly apparent in Fig.~\ref{fig:inside_pump}) between these isolation valves to recuperate the LXe/GXe trapped in the inner volume when a pump is taken out of service and isolated, to purge with a clean dry gas when the volume is opened, and finally to evacuate it before the pump is ready to be brought back in service. Additionally, each pump inlet is equipped with a sintered SS 316L particulate filter to prevent small foreign objects from damaging the pump inducer or impeller.


The volumetric flow at which the efficiency of this pump model is expected to be the highest is
\SI{28}{\literperminute}, much larger than the maximum required flow of \SI{5}{\literperminute} for the
system. A controllable bypass flow (\bypassflow) is therefore fed back at the inlet of the pump through the proportional valve PCV-287 so that the pump can be operated near its peak efficiency flow rate while remaining free to choose which fraction of the flow through the pump (\pumpflow) circulates in the purification loop to the detector (\venturiflow). The pump bypass makes it effectively possible to run the LXe purification loop at a significantly lower flow rate (see Sec.~\ref{subsubsec:perf:flow:pumps}) than the nominal flow rate of the pump, which in general is not possible for cryogenic liquid pumps. This flexibility, however, comes with an added requirement. If no precautions were taken, the temperature change resulting from the enthalpy change across the pump would result in a continuous increase of the LXe temperature, and eventually, the LXe at the inlet would be close to saturated conditions and the pump would cavitate. The \lntwo-LXe heat exchanger HX-288 is installed at the inlet of the pump to eliminate this problem.
\subsection{\lntwo-LXe heat exchanger}
\label{subsec:he}
The \lntwo-LXe heat exchanger (HX-288) used is a welded SS parallel-flow plate heat exchanger custom-designed by DATE~\cite{date} to have a cooling power of \SI{\sim 500}{\watt} at a LXe flow of \SI{15}{\literperminute} and a \lntwo flow of \SI{0.15}{\literperminute}. The \lntwo flow is regulated by a control valve (TCV-288) to about \SIrange[range-phrase=~--~,range-units=single]{0.1}{0.12}{\literperminute} during operation. The heat exchanger is designed such that the thermal contact area between the \lntwo and the LXe is small enough to prevent freezing on the LXe side, as long as the LXe flow is large enough. The HX-288 heat exchanger provides another considerable benefit. It makes it possible to effectively eliminate the heat input into the cryostat that would normally be incurred by the operation of the LXePUR and RRS system. The LXe flowing into the LXePUR system can be subcooled in such a manner as to make its return temperature essentially the same or even lower than that of the LXe of the cryostat. In the latter case the LXePUR system becomes a net-negative heat input to the cryostat. Short timescale temperature fluctuations in the temperature control loop (TIC-288, TCV-288) are damped by the time constant of the LXe purification loop.
\subsection{Purifiers}
\label{subsec:purifiers}
The two redundant purifier units of the LXePUR system are at the core of the system since this is where impurities are removed from the LXe. As mentioned earlier, each LXe purifier unit is composed of a packed sorbent bed designed to remove \otwo, a purifier vessel (F-271/274), and a particulate filter (F-272/275), downstream of the purifier vessel~(see Fig.~\ref{fig:pid}). The design was chosen to enable the use of multiple types of purifier vessels with a range of different geometries. Each purifier unit can be isolated by pneumatic inlet (FV-277/FV-279) and outlet (FV-278/280) valves. Furthermore, internal, manual, and secondary isolation valves (HV-271/275 and HV-272/276) are provided to eliminate the risk of introduction of impurities during purifier replacement operations or in-situ regeneration operations.  The purifier vessels are equipped with SS sheathed cable heaters (R-275/277) for the latter. Additional valves (HV-292/293 and FV-294/295) allow evacuation or inert/reactive gas purge of a purifier unit while its purifier vessel remains installed.
\begin{figure}[!h]
    \centering
	\includegraphics[width=0.45\linewidth]{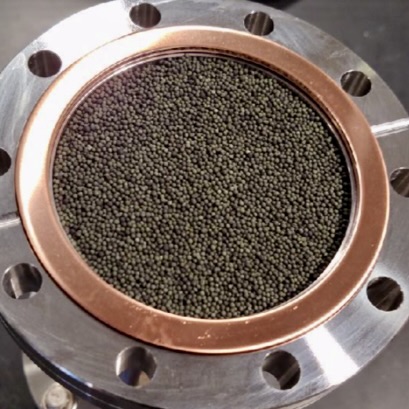}
 	\includegraphics[width=0.45\linewidth]{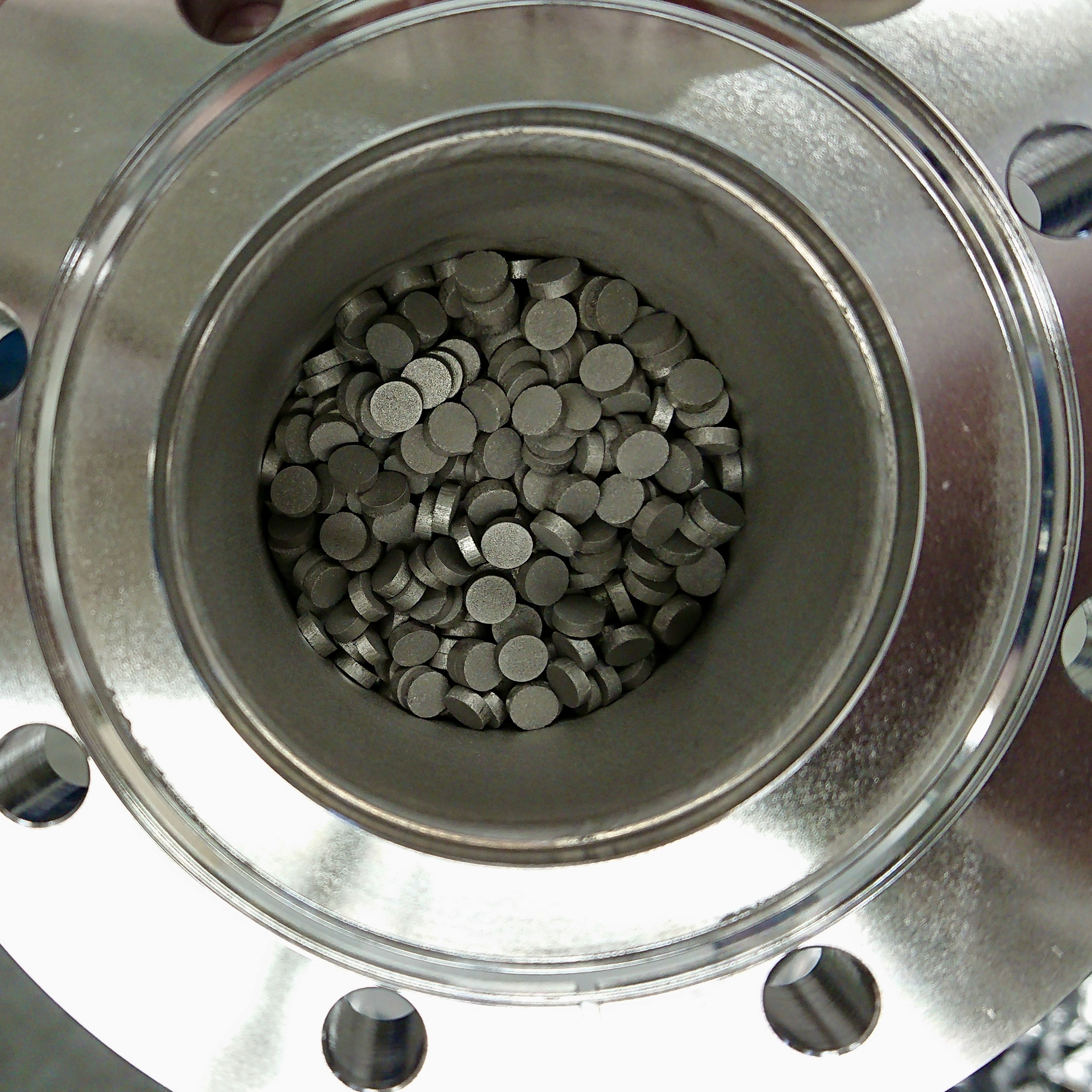}
    \caption{\textbf{Left:} The Q5 material is made from a copper catalyst on high-surface-area alumina support. \textbf{Right:} The St707 material is based on sintered pellets of non-evaporable getter alloy.}
    \label{fig:material}
\end{figure}

Each purifier unit has an independent vacuum-insulated enclosure that houses the LXe purifier and the particulate filter vessels (see Fig.~\ref{fig:inside_purifier}). Each enclosure has an internal diameter of \SI{\sim 40}{\centi\meter} and a height of \SI{1.27}{\meter}, which translates into a usable cylindrical volume with a \SI{\sim 15}{\centi\meter} diameter and a \SI{\sim 1}{\meter} height. The additional space is taken by the particulate filter vessel and the interconnecting piping. All LXe-carrying piping and vessels are covered with MLI.

The particulate filter used is a \SI{2}{\inch} OD sintered SS 316L porous tube manufactured by Boegger (Boedon)~\cite{boedon} with a total surface area of \SI{\sim 800}{\centi\meter\squared} and a nominal media filtration grade of \SI{0.5}{\micro\meter}. The porous tube is itself enclosed in a \SI{3}{\inch} OD SS cylindrical vessel. The LXe flows from the outside of the porous tube to the inside, and then exits the particulate filter vessel through the bottom. The large surface area of the porous tube was chosen to make the pressure drop at the maximum required flow subdominant, leaving a larger margin available for the LXe purifier vessel.

Two LXe purifier types have been chosen and designed based on the test results described in~\cite{xeclipse}, as well as from tests with prototype purifiers (Sec.~\ref{subsubsection:perf:pur:prototypes}). Both purifier types were designed such that their pressure drop would be compatible with the expected delivered pressure of the cryogenic liquid pump (at its maximum frequency) at the maximum flow rate required.
\begin{table*}[!h]
\caption{Characteristics of the two purifier types used in the XENONnT LXePUR system.}
\centering
\begin{tabular}{ccccc}
\hline
 Purifier & Material & Pellet size [mm] & Sorbent bed [mm] & Mass [g] \\ \hline
 1 & Q5 & $\diameter$\SIrange[range-phrase=~--~,range-units=single]{0.6}{1.2}{} & $\diameter47.5\times341$ & 475 \\
 2 & St707 &$\diameter6\times2$& $\diameter57\times510$&  3000 \\ \hline
\end{tabular}
\label{tab:purifier_types}
\end{table*}

The first LXe purifier type (``Purifier 1'', Table~\ref{tab:purifier_types}) has a large sorption capacity for \otwo compared to the expected total initial amount of impurities present in the LXe before starting the purification system. The sorbent used for this purifier is the catalyst Engelhard Q5 (also known as BASF Cu-0226S), consisting of copper-impregnated spherical alumina pellets with a diameter of \SIrange[range-phrase=~--~,range-units=single]{0.6}{1.2}{\milli\meter} ($14\times28$ mesh) that can adsorb \otwo through oxidation of the copper (see Fig.~\ref{fig:material}). Once oxidized, the sorbent pellets can be regenerated by flowing a hydrogen (5\%) argon (95\%) gas mixture through the bed at \SI{\sim 200}{\celsius}. The sorbent bed of the purifier has a diameter of \SI{47.5}{\milli\meter} and a length of \SI{341}{\milli\meter} for a sorbent mass of \SI{\sim 475}{\gram}. It is contained within a double-flanged (CF) cylindrical SS vessel with an inlet/outlet CF-to-VCR adapter flanges equipped with \SI{100}{\micro\meter} filtration grade sintered SS 316L porous discs (flow area \SI{\sim 20}{\centi\meter\squared}). The \rnbkg emanation from this purifier could be as high as \SI{180}{\milli\becquerel} \cite{xenonnt:radiopurity}, significantly exceeding the stringent experimental requirement of less than \SI{1}{\milli\becquerel} per purifier. This precludes its long-term use. However, its \otwo removal efficiency was expected to be more than 90\% and its sorption capacity greater than \SI{100}{\milli\gram} without any deterioration in efficiency.

The second LXe purifier type  (``Purifier 2'', Table~\ref{tab:purifier_types}) was designed to have a much more limited \otwo sorption capacity (estimated to be at least a few mg) and a relatively smaller \otwo removal efficiency (estimated to be \textgreater 50\%) but with a \rnbkg emanation rate of \SI{0.24}{\milli\becquerel}~\cite{xenonnt:radiopurity}, compatible with the stringent requirements. The sorbent chosen for this purifier is the SAES St707 alloy of zirconium, vanadium, and iron, in the form of cylindrical pellets with \SI{6}{\milli\meter} diameter and \SI{2}{mm} height (see Fig.~\ref{fig:material}). The sorbent bed of the purifier has a diameter of \SI{\sim 57}{\milli\meter} and a length of \SI{\sim 510}{\milli\meter}, for a total mass of St707 of \SI{3}{\kilogram}. It is contained in a cylindrical SS vessel with an inlet \SI{40}{\micro\meter} filtration grade sintered SS 316L porous disc and an outlet \SI{3}{\micro\meter} grade porous cup (Fig.~\ref{fig:inside_purifier}). The St707 is activated by heating the purifier vessel to \SI{\sim 350}{\celsius} for 24 hours under vacuum. The purifier can be regenerated in-situ, either under vacuum, or in a GXe atmosphere. Selected results from the purifiers are described in Sec.~\ref{sec:performance}.
\subsection{Purity monitor}
\label{subsec:pm}
\begin{figure*}[!h]
    \centering
    \includegraphics[width=\linewidth]{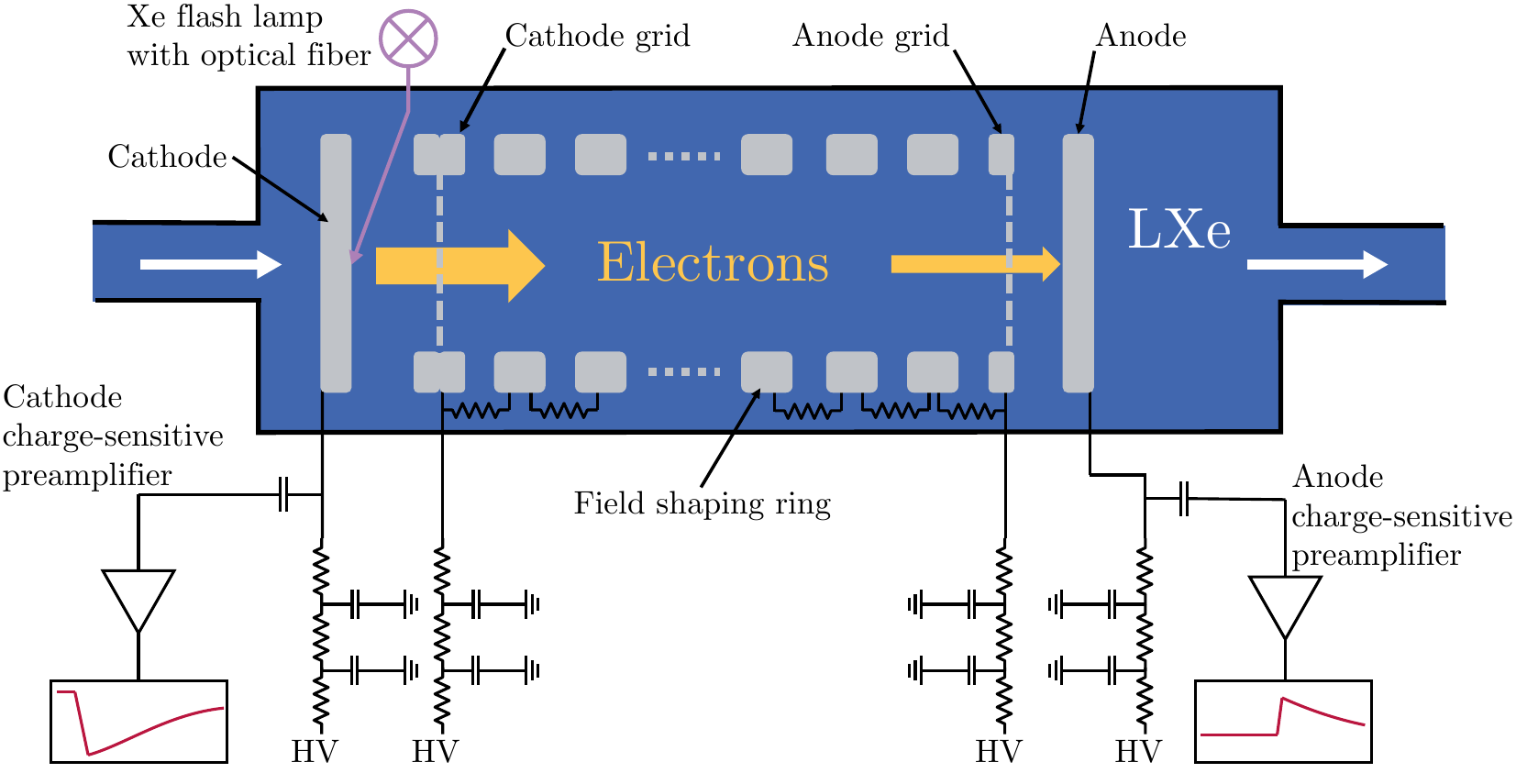}
    \caption{Schematic drawing of the purity monitor. Electrons are extracted from the photocathode using a xenon flash lamp coupled to a UV quartz optical fiber. The cathode and anode are connected to charge-sensitive preamplifiers individually. The cathode and anode signal are recorded by the waveform digitizer. The cathode, cathode grid, anode grid, and anode are biased to \SIlist{-50;0;2040;2540}{\volt}, respectively. A set of 16 field shaping rings, positioned between the cathode and anode grids and biased at intermediate potentials via a voltage divider network, keep the drift field uniform. Liquid xenon (LXe) is continuously flowing through the PM from left to right during operation.}
    \label{fig:pm_description}
\end{figure*}

A purity monitor (PM) was developed to constantly monitor and directly measure the purity level achieved by the LXePUR system. Its design is based on the same principle as the device developed for this purpose by the ICARUS collaboration~\cite{icarus:pm}. PMs have since been deployed and their long-term use was demonstrated in multiple liquid argon experiments ~\cite{icarus:purity,dunecollaboration2018dune}. The PM is a double-grid ionization chamber with four electrodes: a cathode disk with a photocathode at its center, a cathode grid, an anode grid, and an anode disk. The purity of the liquid is obtained by comparing the charge leaving the cathode, $Q_\mathrm{C}$, extracted via the photoelectric effect, with the charge collected at the anode, $Q_\mathrm{A}$, after its drift between the two electrodes and attenuation due to attachment to impurities. The electron lifetime, $\tau_e$, can be directly calculated following Ref.~\cite{icarus:pm} via
\begin{linenomath*}
\begin{equation}
    \tau_e = -\frac{t_\mathrm{D}}{\ln \left({Q_\mathrm{A}/Q_\mathrm{C}}\right)},
 \label{eq:pm_elife}
\end{equation}
\end{linenomath*}
where $t_\mathrm{D}$ is the drift time of the electrons between the two grids. The design of the PM of the LXePUR system, shown in Fig.~\ref{fig:pm_description}, shares many characteristics with the one built for the Xeclipse facility~\cite{xeclipse}. The cathode, both grids and anode are made of stainless steel, with a diameter of \SI{50}{mm}. The cathode disk center is gold-plated with a diameter of \SI{12}{mm} doubling as photocathode. Different to Ref.~\cite{xeclipse}, both grids feature a hexagonal shape with a line width of \SI{0.1}{mm} and a distance between parallel lines of \SI{2}{mm} instead of a wire structure. More details are described in Ref.~\cite{nobuthesis}.

Electrons are extracted from the photocathode using a Hamamatsu L7685 xenon flash lamp~\cite{hamamatsu} coupled to a \SI{600}{\micro\meter} core UV quartz optical fiber. The cathode (anode) charge is measured by integrating the current induced on the electrode while the electrons drift to (from) the corresponding grid electrode with Amptek A250 charge-sensitive preamplifiers~\cite{amptek}. The cathode and anode signal waveforms are recorded on separate channels by a CAEN DT5724 14\,bit 100\,MS/s digitizer~\cite{caen}. The cathode, cathode grid, anode grid, and anode are biased to \SIlist{-50;0;2040;2540}{\volt}, respectively, and are separated by \SIlist{17;204;9}{\milli\meter}. Thus, the electric field between the cathode grid and the anode grid is \SI{100}{\volt\per\centi\meter}. A set of 16 field shaping rings (\SI{38}{mm} inner diameter, \SI{10}{mm} height) positioned between the cathode and anode grids and biased at intermediate potentials via a voltage divider network, keep the drift field uniform. The electric fields are established to ensure that the electron transmission efficiency of both grids is \SI{\sim 100}{\percent} verified with electric field simulations using COMSOL.

The PM is installed in the LXe purification loop flow path on the LXePUR skid outside of the water tank (see Figs.~\ref{fig:pid} and \ref{fig:loop}). This location provides two significant advantages over installation in the cryostat vessel: maintenance or repairs can be conducted when needed, with minimal disruption to detector operation, and it enables the possibility of measuring the concentration of impurities at different locations along the LXe flow path of the skid. Specifically, the PM is located such that measurements upstream (FV-275, FV-283, FV-276) or downstream (FV-281, FV-284, FV-282) of the purifier units are possible. The former is representative of the purity in the detector, while the latter can be used to verify the proper operation of the purifiers. These two sampling modes have two independent arrangements of isolation/bypass valves, also allowing one to isolate the PM vessel entirely from the purification loop for maintenance. It is also possible to bypass both purifier units, provided that flow through or around the PM vessel is maintained.

Typically, an electron lifetime measurement is performed every \SI{15}{\minute} allowing to monitor and model the purity evolution over time, and to enlarge the service life of the flash lamp. To reduce the effect of baseline fluctuations, the flash lamp is pulsed at a frequency of \SI{10}{\hertz} for \SI{30}{\second}, and an average of 300 waveforms is used for each charge signal. Since the measurement principle relies on the survival probability of electrons over the drift distance, the accuracy of electron lifetime measurements is limited by the systematic uncertainty of the signal ratio ${Q_\mathrm{A}/Q_\mathrm{C}}$. The stability and gain difference of the two preamplifiers, therefore, contribute to the systematic uncertainty. The systematic uncertainty is evaluated using a test pulse calibration of the preamplifiers. The overall systematic uncertainty on the charge signal ratio is approximately \SI{0.3}{\percent}~\cite{nobuthesis}. It translates to a relative uncertainty on electron lifetime measurements of \SI{10}{\percent} at \SI{5}{ms} and \SI{20}{\percent} at \SI{10}{ms}. The PM can therefore provide precise measurements (\SI{\lesssim 20}{\percent} relative uncertainty) of \otwo-equivalent concentrations in LXe down to \SI{\sim 29}{\ppt} (Eq.~(\ref{eq:elconversion})).
\section{System performance}
\label{sec:performance}
This section synthesizes the results obtained to assess the performance of the LXePUR system in three areas of interest: achievable LXe flow and resulting resistances (Sec.~\ref{subsec:perf:flow}), heat loads (Sec.~\ref{subsec:perf:therm}), and purification (Sec.~\ref{subsec:perf:pur}). A number of results were obtained in two commissioning periods, September-October 2019, and January-February 2020, prior to the installation of the XENONnT TPC in the cryostat. The first commissioning test sequence focused on: validating the LXe pump start-up procedure, verifying proper continuous operation of the PM over many days, evaluating the performance of the LXe pumps, comparing expected pressure losses along the purification loop with direct measurements, and finally testing purification of the LXe volume filled into the cryostat with prototype Q5 purifier vessels. The second commissioning test sequence involved a detailed characterization of the LXe pump performance and an analysis of the flow and purification performance of additional prototype Q5 purifiers and a prototype St707 purifier. The main purification performance results were obtained with long term operation of the system during the commissioning phase of the experiment with the TPC and during its first science run~\cite{PhysRevLett.129.161805}.
\subsection{Flow performance and pressure losses}
\label{subsec:perf:flow}
The equilibrium LXe purity level and the time required to reach it are directly determined by the achievable LXe flow rate through the purification loop and the efficiency of the purifier (Eq.~(\ref{eq:equilibium_purity})). It is therefore important to verify that system components perform according to expectations in order to obtain a desired LXe purity. In addition, the quantitative characterization of components in terms of pressure generation, in the case of the LXe pumps, or flow resistance enables the development of a fluid flow model that can predict conditions along the purification loop and help choose specific operating conditions.

As with any flowing cryogenic liquid, the LXe fluid state can change substantially along the flow path of the purification loop. In order to achieve the most stable flow conditions, LXe should be in a subcooled state along most of the flow path, except in the portion of the LXe return line after its highest point (Fig.~\ref{fig:loop}), which has a negative slope. Open channel flow conditions are prevalent in that portion of the LXe return line since the outlet is opened to the cryostat vessel GXe volume. From the LXe surface in the cryostat vessel, through the vessel bottom port, and down to the lowest point of the inlet line to the LXePUR skid (Fig.~\ref{fig:loop}), the LXe fluid pressure will increase substantially (\SI{\sim 1.5}{\bar}) due to the \SI{5.4}{\meter} vertical drop. However, since the heat input along the piping is low compared to the fluid velocity (Sec.~\ref{subsubsec:perf:therm:piping}), the LXe state becomes progressively subcooled. The \lntwo-LXe heat exchanger (HX-288) further subcools the LXe before it enters the LXe pump. The LXe pump produces a pressure increase \pumpdp and also increases the LXe temperature.

In normal operation, a pressure differential (\pumpdp) of \SI{\sim 1.7}{\bar} is required to drive a flow (\venturiflow) of \SI{2.4}{\literperminute} through the LXe purification loop, which includes a flow of \SI{\sim 0.4}{\literperminute} through the radon removal system. The majority of components of the LXe purification loop have been sized to have a low flow resistance so that the flow performance of the system would be minimally impacted by their presence. For example, the \lntwo-LXe heat exchanger (HX-288) was designed to have a pressure drop smaller than \SI{50}{\milli \bar} at \SI{15}{\literperminute}. Similarly, the flow coefficient chosen for the on/off valves of the system (FV-2XX) results in a pressure drop per valve of \SI{5}{\milli \bar} at \SI{2.4}{\literperminute}. Also, the filtration surface, media grade, and thickness of the particulate filter (F-272/275) were chosen such that the resulting pressure drop at \SI{2.4}{\literperminute} would be below \SI{40}{\milli \bar}. The largest pressure drops are across the purifier vessel, \SI{\sim 400}{\milli \bar} for Purifier 2, across the proportional valve FCV-286 that diverts a fraction of the LXe flow to the radon removal system, \SI{\sim 450}{\milli \bar} (valve opened at 10\%), and across the Venturi (FE-288), \SI{\sim 60}{\milli \bar}, all at \SI{2.4}{\literperminute}. The remaining differential pressure, \SI{\sim 680}{\milli \bar}, is required to compensate for the hydrostatic pressure of the LXe column from the LXe level in the TPC to the highest point of the LXe return line (Fig.~\ref{fig:loop}) and other minor piping losses. The coarse media grade of the LXe pump inlet particulate filter and the large unobstructed cross-sectional area of the PM vessel both result in a negligible flow resistance.

The tests performed to characterize the pressure-flow performance of the two LXe pumps (Sec.~\ref{subsubsec:perf:flow:pumps}), the frictional pressure losses in the purifiers (Sec.~\ref{subsubsec:perf:flow:purifiers}), and in the Venturi tube (Sec.~\ref{subsubsec:perf:flow:venturi}), are described below.
\begin{figure}[t]
    \centering
    \includegraphics[width=\columnwidth]{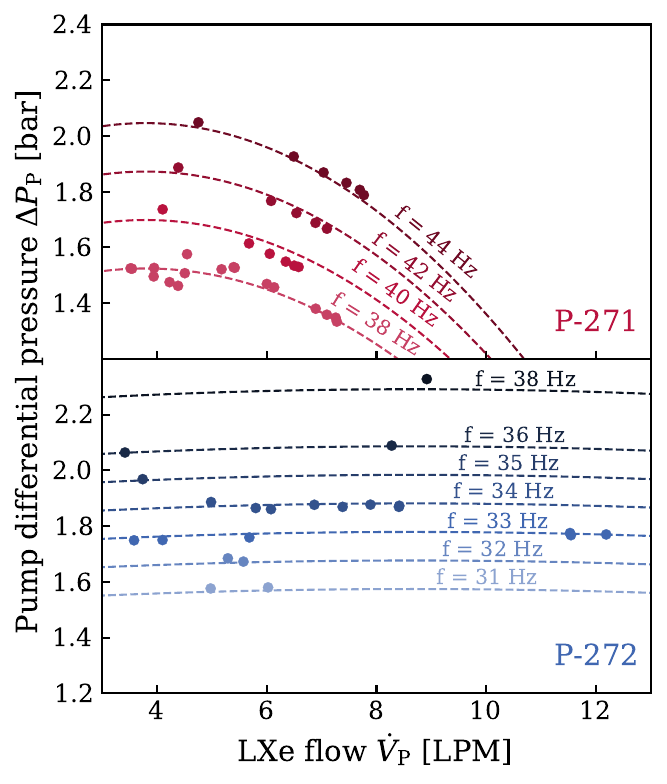}
    \caption{Performance of LXe pumps P-271 (top, red) and P-272 (bottom, blue) in terms of differential pressure \pumpdp as a function of the pump flow rate \pumpflow for several values of the pump motor driving frequency $f$. The pump differential pressure and flow rate for P-271 (P-272) are calculated using the measured pressures PT-271 and PT-275 (PT-276), the purification loop flow rate \venturiflow, and the expected pressure drop across valves PCV-287, HV-277 (HV-278), and FV-272 (FV-274), using Eqs.~(\ref{eq:pumpflow}) and (\ref{eq:pumpdp}). The dashed lines show a fit of Eq.~(\ref{eq:pump_curve}) to the performance data. The performance from P-272 is compatible with the estimate from the manufacturer with a flat characteristic while P-271 exhibits a decreasing characteristic and requires higher driving frequency to achieve the same differential pressure at a given flow rate. Nevertheless, the performance is sufficient to achieve the required purification loop flow of the experiment.}
    \label{fig:pump_curves}
\end{figure}
\subsubsection{LXe pump characteristic curves}
\label{subsubsec:perf:flow:pumps}
The established approach for describing hydraulic pump performance is via the determination of the characteristic curves of a pump~\cite{pumphandbook}, that is, the differential pressure (or equivalently the pressure head) produced by the pump as a function of flow rate. In the case of the LXePUR system, the lack of a measurement of the pump flow rate (\pumpflow), and to a lesser extent the fact that each pump has a valve (HV-277 and HV-278 for P-271 and P-272, respectively) upstream before the closest pressure transducer (PT-271), implies that it is not possible to measure the characteristic curves of the LXe pumps directly. It is possible, however, to calculate the pump flow rate (\pumpflow) and the differential pressure produced by the pump (\pumpdp) using the purification loop flow measurement (\venturiflow) and the knowledge of the flow resistances in the pump bypass loop. In the case of P-271, starting from pressure transducer PT-275 and going along the pump bypass flow path, neglecting the pressure losses due to the piping, the pressure decreases according to
\begin{linenomath*}
\begin{equation} \label{eq:bypass_pdec}
	\pressureat{PT}{275} - \deltapof{FV}{272} - \deltapof{PCV}{287} - \deltapof{HX}{288} =
	\pressureat{PT}{271}. \\
\end{equation}
\end{linenomath*}
The pressure drop across a valve can be expressed in terms of the flow coefficient \flowcoefficient of the valve as~\cite{valvehandbook}
\begin{linenomath*}
\begin{equation} \label{eq:fv_cv}
	\Delta P = \rho_r \frac{\dot{V}^2}{\flowcoefficient^2},
\end{equation}
\end{linenomath*}
where $\rho_r$ is the relative density of the fluid with respect to that of water, $\dot{V}$ the volumetric flow, with both $\Delta P$ and $\dot{V}$ expressed in U.S. customary units. Using Eq.~(\ref{eq:fv_cv}), and the fact that the pressure drop across HX-288 is expected to be significantly smaller than that across even FV-272, Eq.~(\ref{eq:bypass_pdec}) becomes
\begin{linenomath*}
\begin{equation}
  \frac{\pressureat{PT}{275}-\pressureat{PT}{271}}{\rho_r} =  \frac{\pumpflow^2}{\cvof{FV}{272}^2} +
         \frac{\bigl({\pumpflow-\venturiflow}\bigr)^2}{\yof{PCV}{287}^2 \cvof{PCV}{287}^2},
\end{equation}
\end{linenomath*}
where \yof{PCV}{287} is the fractional opening of valve PCV-287, and where the flow through the valve was expressed in terms of the pump flow rate and the purification loop flow rate with $\pumpflow = \bypassflow + \venturiflow$. Hence the pump flow rate can be calculated via
\begin{linenomath*}
\begin{equation} \label{eq:pumpflow}
	\pumpflow = \phi \venturiflow + \sqrt{(1-\phi) \bigl({\dot{V}_{\mathrm{virt}}^2 - \phi \venturiflow^2}\bigr)},
\end{equation}
\end{linenomath*}
where
\begin{linenomath*}
\begin{equation} \label{eq:phi}
	\phi \equiv \cvof{FV}{272}^2/\left({\cvof{FV}{272}^2+\yof{PCV}{287}^2 \cvof{PCV}{287}^2}\right),
\end{equation}
\end{linenomath*}
and
\begin{linenomath*}
\begin{equation} \label{eq:virtflow}
    \dot{V}_{\mathrm{virt}}^2 \equiv \cvof{FV}{272}^2 \left({\pressureat{PT}{275}
    -\pressureat{PT}{271}}\right) / \rho_r.
\end{equation}
\end{linenomath*}
Similarly, but this time starting from pressure transducer PT-271 and following the flow path through the pump, the pressure varies according to
\begin{linenomath*}
\begin{equation}
	\pressureat{PT}{271} - \deltapof{HV}{277} + \pumpdp = \pressureat{PT}{275}.
\end{equation}
\end{linenomath*}
Hence, using the fact that $\cvof{HV}{277} = \cvof{FV}{272}$, the differential pressure produced by the pump can be calculated from
\begin{linenomath*}
\begin{equation} \label{eq:pumpdp}
    \pumpdp = \rho_r \bigl({\dot{V}_{\mathrm{virt}}^2 + \pumpflow^2}\bigr)/\cvof{HV}{277}^2,
\end{equation}
\end{linenomath*}
with $\pumpflow$ and $\dot{V}_{\mathrm{virt}}$ given by Eqs.~\ref{eq:pumpflow} and \ref{eq:virtflow}. Note that the pump flow rate and the differential pressure produced by P-272 can also be calculated from Eqs.~\ref{eq:pumpflow} and \ref{eq:pumpdp} simply by letting $\pressureat{PT}{275} \rightarrow \pressureat{PT}{276}$ in the expression for $\dot{V}_{\mathrm{virt}}^2$ (Eq.~(\ref{eq:virtflow})) since $\cvof{FV}{274} = \cvof{HV}{278} = \cvof{FV}{272}$.

During the second commissioning test sequence of the LXePUR system, both LXe pumps were operated over a range of driving frequencies and flows in order to determine their characteristic curves. The pump flow rate was varied over a wide range by changing the purification loop flow rate (\venturiflow), via FCV-286 openings between 6\% and 65\%, and by changing the bypass flow rate (\bypassflow), via PCV-287 openings between 7\% and 45\%, as can be deduced from an examination of Eq.~(\ref{eq:pumpflow}). These commissioning tests were conducted with LXe pump inlet pressures (PT-271) of \SI{\sim 3.0}{\bar} due to the lower liquid level in the cryostat vessel compared to the TPC liquid level in normal operation. Figure~\ref{fig:pump_curves} shows the performance of both P-271 (red points) and P-272 (blue points) in terms of differential pressure and flow rate for different driving frequencies, as calculated by Eqs.~\ref{eq:pumpflow} and \ref{eq:pumpdp}. The characteristic curves of the LXe pumps are modeled with
\begin{linenomath*}
\begin{equation} \label{eq:pump_curve}
	\pumpdp \! \bigl({\pumpflow, f}\bigr) = \bigl({a_0 + a_1 f}\bigr) 
	\bigl({b_0 + b_1 \pumpflow +b_2 \pumpflow^2}\bigr) + c_0, 
\end{equation}
\end{linenomath*}
where $f$ is the pump driving frequency and the parameters $\{a-c\}_i$ are associated with a given pump. The form of Eq.~(\ref{eq:pump_curve}) was chosen so that the differential pressure produced by the pump would be linear with the driving frequency, and quadratic with the flow rate in order to be able to represent a flat or a gradually falling characteristic~\cite{pumphandbook}. The dashed lines in Fig.~\ref{fig:pump_curves} show the fit to Eq.~(\ref{eq:pump_curve}) for each pump. The parameters $\{a-c\}_i$ obtained are listed in Tab.~\ref{tab:pump_performances}. The performance for P-272 is close to the estimated performance, a mostly flat characteristic with \SI{\sim 2}{\bar} differential pressure at \SI{12}{\literperminute} and \SI{34}{\hertz}, from the pump manufacturer~\cite{barbernichols}. In contrast, P-271 exhibits a decreasing characteristic and requires higher driving frequency to achieve the same differential pressure at a given flow rate. This could be due, for example, to a slightly greater gap between the pump impeller blades and the casing. The differential pressure (\pumpdp) of \SI{\sim 1.7}{\bar} required to drive a flow (\venturiflow) of \SI{2.4}{\literperminute} through the purification loop in normal operation (Sec.~\ref{subsec:perf:flow}) can be accommodated with a bypass valve PCV-287 opening of 10\% and a driving frequency of \SI{\sim 40}{\hertz} (P-271), or \SI{\sim 32}{\hertz} (P-272), resulting in pump flow rates \pumpflow of \SI{\sim 5}{\literperminute}. The maximum pump motor frequency is given as \SI{50}{\hertz} by the manufacturer. Driving frequencies in excess of \SI{\sim 42}{\hertz} are expected to lead to greater wear of the cold bearing. Differential pressures in excess of \SI{\sim 1.9}{\bar} are therefore not easily accessible with both LXe pumps. A model such as Eq.~(\ref{eq:pump_curve}) to quantitatively describe the performance of the LXe pumps is helpful. It can be used, along with the flow resistances of the other components of the system, for example, to predict the performance in conditions outside of those tested, or to choose in advance approximate operating conditions such as pump driving frequency and control valve openings to achieve a specific purification loop flow rate.

\begin{table*}[!h]
\caption{Best-fit parameters for the pump differential pressure as a function of flow and driving frequency (Eq.~(\ref{eq:pump_curve})).}
\centering
\begin{tabular}{ccccccc}
\hline
 Pump & $a_0$ & $a_1$ & $b_0\times 10^{2}$ & $b_1 \times 10^{3}$ & $b_2 \times 10^{4}$ & $c_0$ \\
 & $[\unit{\bar}]$ & $[\unit[per-mode=symbol]{\bar\per\hertz}]$ & $[1]$& $[\unit{\per\literperminute}]$ & $[\unit{\per\square\literperminute}]$ & $[\unit{\bar}]$ \\ \hline
 P-271 & $0.16$ & $1.92$ & $4.22$ & $1.60$ & $-2.11$ & $-1.79$ \\
 P-272 & $0.48$ & $0.63$ & $16.1$ & $0.64$ & $-0.37$ & $-1.68$ \\ \hline
\end{tabular}
\label{tab:pump_performances}
\end{table*}
\subsubsection{Purifier pressure losses}
\label{subsubsec:perf:flow:purifiers}
The sorbent beds contained in the purifier vessels of the LXePUR system must balance the mass of sorbent material against their radiopurity, their \otwo removal efficiency and sorption capacity, and their resistance to LXe flow. While the pressure drop can be predicted at the 10\%-20\% level from a given purifier vessel geometry and sorbent material geometry, it is not so easily done for the flow dependence of the \otwo removal efficiency. This is mostly due to the uncertainty in extrapolating the results obtained in \cite{xeclipse} by more than an order of magnitude in flow rate (\SI{\sim 0.2}{\literperminute} $\rightarrow$ \SI{\sim 5}{\literperminute}). Furthermore, since the results on the Q5 sorbent in \cite{xeclipse} indicated an \otwo removal efficiency at \SI{0.15}{\literperminute} close to 100\% for a relatively small mass of sorbent (\SI{18}{\gram}), it was not initially clear which sorbent mass, sorbent type, and purifier geometry would provide the best balance. For these reasons, a number of LXe purifier types, of various aspect ratios and sorbent types were evaluated during the commissioning tests of the LXePUR system before the installation of the TPC as summarized in Tab.~\ref{tab:purifier_types_proto}.
\begin{table*}[t]
\caption{Prototype purifiers for the commissioning of the LXePUR system, enclosed by porous sintered discs or cups. Purifier C1 (Q5) and C3 (St707) were equipped with inlet/outlet Swagelok SS-6BW-TW ($\flowcoefficient=1$) and SS-8BW-TW ($\flowcoefficient=1.1$) valves, respectively, to prevent accidental sorbent passivation from air exposure during installation. Note that the St707 pellets in prototype C3 had a slightly different geometry than those used in the final Purifier 2.}
\centering
\begin{tabular}{ccccccc}
\hline
Purifier & Material & Pellet size [mm] & Sorbent bed [mm] & Mass [g] & Inlet/outlet filter & Rating [\si{\micro\meter}] \\ \hline
C1 & Q5 & $\diameter$\SIrange[range-phrase=~--~,range-units=single]{0.6}{1.2}{}&$\diameter28.6\times20.4$& 10.7 & Disc & 100 \\ 
C2 & Q5 & $\diameter$\SIrange[range-phrase=~--~,range-units=single]{0.6}{1.2}{} &$\diameter22.1\times190$& 61.9 & Cup & 90  \\ 
C3 & St707 & $\diameter4\times2$ & $\diameter97.4\times122$& 2400 & Disc & 0.5 \\ \hline
\end{tabular}
\label{tab:purifier_types_proto}
\end{table*}

The pressure loss from a purifier unit can be calculated simply by subtracting the downstream pressure measurement (PT-286) from the corresponding upstream pressure measurement, PT-273 if the purifier is installed as F-271, or PT-274 if it is installed as F-274. As mentioned earlier, the flow resistance of the particulate filters (F-272/275) is expected (by design) to be significantly smaller than that of the packed sorbent beds while the flow resistance of the purifier unit outlet valves (HV-272 and FV-278, or HV-276 and FV-280) is essentially negligible. 

\begin{figure}[!h]
    \centering
    \includegraphics[width=\columnwidth]{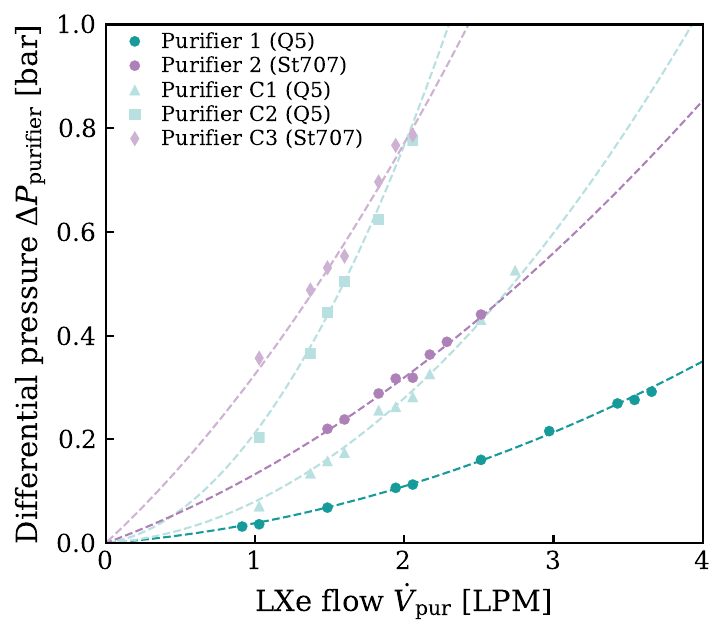}    
    \caption{Pressure drop across Purifier 1 (Q5, teal) and Purifier 2 (St707, violet) of the XENONnT LXePUR system as a function of LXe flow \venturiflow, as well as for prototype purifiers C1 (Q5, triangles), C2 (Q5, squares), and C3 (St707, diamonds). The data is modeled using Eq.~(\ref{eq:purifier_dp}) (dashed lines) with best-fit parameters listed in Tab.~\ref{tab:purifier_dp}.}
    \label{fig:purifier_dp}
\end{figure}

Figure~\ref{fig:purifier_dp} shows the pressure loss of Purifier 1 (Q5) and Purifier 2 (St707) as a function of flow rate, as well as the loss for the prototype purifiers C1, C2, and C3. The quadratic dependence on the flow rate is apparent for all purifiers, indicating that with the pellet dimensions, flow rates, and geometries considered here, the nature of the flow in the purifiers is turbulent. The pressure loss of Purifier 1 is significantly smaller than that of Purifier 2, mostly due to the finer filtration grade of the porous cup installed at the outlet of the latter (Sec.~\ref{subsec:purifiers}). A large fraction of the pressure losses of Purifier C1 ($\sim$40\% at \SI{2}{\literperminute}) is caused by its inlet/outlet valves. For Purifier C3 the effect of its valves is smaller ($\sim$12\%) and the pressure losses are dominated ($\gtrsim$80\%) by its \SI{0.5}{\micro\meter} sintered porous discs. The pressure losses for Purifier C2 are mainly due to its packed bed having a small cross section for flow and a long depth.

The Ergun equation~\cite{ergun} is commonly used to describe the pressure drop across a packed bed. It combines both a laminar term, proportional to the fluid viscosity and the fluid velocity, and a turbulent term, proportional to the fluid density and quadratic with the fluid velocity. Accordingly, the flow dependence of the pressure loss is modeled with
\begin{linenomath*}
\begin{equation} \label{eq:purifier_dp}
	\Delta P_\mathrm{purifier} = k_1 \venturiflow + k_2 \venturiflow^2,
\end{equation}
\end{linenomath*}
where $k_1$ and $k_2$ are parameters that are specific to a given purifier. Figure~\ref{fig:purifier_dp} shows the resulting fits to the measured pressure losses (dashed lines) and the values of the parameters are listed in Tab.~\ref{tab:purifier_dp}. The values of $k_1$ and $k_2$ obtained for Purifier 1 (Q5) are compatible with the prediction from the Ergun equation for a randomly packed bed ($\sim$36\% porosity) with spherical sorbent pellets with a diameter of \SI{1.2}{\milli\meter}. As discussed above, the larger $k_1$ values for some purifiers is indicative of the finer grade sintered porous filters used in their construction since such components are akin to a multitude of parallel capillaries with Hagen-Poiseuille flow, and hence, with pressure losses proportional to flow velocity.

\begin{table}[!h]
\caption{Best-fit parameters for the pressure loss across a purifier as a function of flow (Eq.~(\ref{eq:purifier_dp})).}
\centering
\begin{tabular}{cccc}
\hline
 Purifier & Material & $k_1$ & $k_2$ \\
 & & $[\unit[per-mode=symbol]{\milli\bar\per\literperminute}]$ & $[\unit[per-mode=symbol]{\milli\bar\per\square\literperminute}]$ \\ \hline
 1 & Q5 & $22$ & $16$ \\
 2 & St707 & $106$ & $27$ \\ \hline
 C1 & Q5 & $20$ & $60$ \\
 C2 & Q5 & $42$ & $171$ \\
 C3 & St707 & $265$ & $60$ \\ \hline
\end{tabular}
\label{tab:purifier_dp}
\end{table}
\subsubsection{Venturi pressure loss}
\label{subsubsec:perf:flow:venturi}
Another non-negligible flow resistance in the purification loop, especially close to the maximum achievable flow, is the Venturi tube (FE-288). Since the Venturi tube measurement principle relies on a pipe segment of reduced cross-sectional area, a pressure drop is expected across the instrument. Note that the pressure drop being referred to here is not the differential pressure used for the flow measurement, between the pressure upstream and in the restriction, but rather the overall pressure drop.  The latter is smaller since there is pressure recovery in the diffuser downstream of the restriction~\cite{nakayama}.

The flow dependence of the overall pressure drop across FE-288 was evaluated during the second commissioning period of the LXePUR system. It was obtained by taking the difference between the pressure measured immediately upstream of FE-288 (PT-286) and the first pressure measured downstream (PT-289), correcting the downstream pressure for its decrease due to the \SI{1.7}{\meter} height difference between the two measurement points, in a wide range of flow conditions.

\begin{figure}[!t]
    \centering
    \includegraphics[width=\columnwidth]{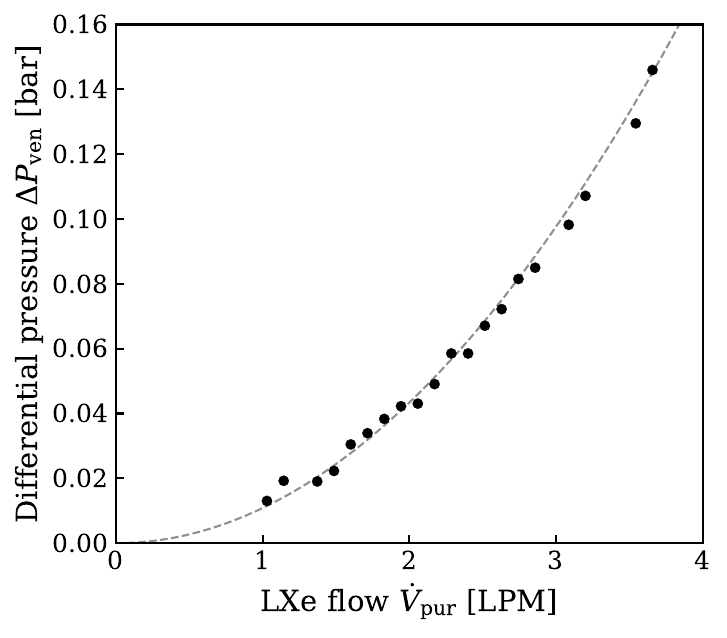}    
    \caption{Pressure drop $\venturidp$ across the Venturi tube (FE-288) as a function of purification loop flow \venturiflow. The Venturi tube relies on a flow area restriction for its measurement thereby contributing to the overall pressure losses of the purification loop. The flow dependence is well described by a quadratic function (dashed line).}
    \label{fig:venturi_dp}
\end{figure}

Figure~\ref{fig:venturi_dp} shows the resulting average Venturi tube pressure drop $\venturidp$ as a function of the purification loop flow \venturiflow, as well as the result of the fit to a quadratic function of the flow, $\venturidp = k_3 \venturiflow^2$ (dashed line). The fitted constant is $k_3 = \SI[per-mode = symbol]{10.9}{\milli\bar\per\square\literperminute}$. The measured flow dependence of the pressure drop is well described by the quadratic function, as would be expected for the pressure drop across a restriction~\cite{nakayama}. In normal operation, the pressure loss of \SI{\sim 60}{\milli\bar} at \SI{2.4}{\literperminute} is not as significant as that of the LXe purifiers and accounts for only a small fraction ($\sim$4\%) of the overall losses. 

\subsection{Thermal performance}
\label{subsec:perf:therm}
The thermal performance of the LXePUR system directly affects its heat load contribution to the XENONnT cryostat and its cryogenic system~\cite{xenonnt:instrument}. The heat input from the cryogenic LXe pump alone, for example, can easily be in excess of \SI{150}{\watt}~(Sec.~\ref{subsubsec:perf:therm:pump}), which is larger than the available cooling power of the cryogenic system. As described in Sec.~\ref{subsec:he}, the LXePUR system \lntwo-LXe heat exchanger (HX-288) can be used to offset this heat load, which is part of the motivation for its inclusion in the first place. Nevertheless, assessing the thermal performance of the system can provide insights into the available margin of the design cooling power of the heat exchanger, as well as for the design of future cryogenic LXe purification systems. Also, as discussed at the beginning of the previous section (Sec.~\ref{subsec:perf:flow}), the most stable LXe flow conditions will be achieved if the fluid remains in a subcooled state along the purification loop flow path, at least up until its highest point, which requires an understanding of the size of the heat inputs into the system.

Estimates of the heat load from individual components of the system are described in Secs.~\ref{subsubsec:perf:therm:piping} (piping and valves), \ref{subsubsec:perf:therm:purifiers} (purifier units), and \ref{subsubsec:perf:therm:pump} (LXe pumps). Sec.~\ref{subsubsec:perf:therm:min_subcooling} discusses the interplay between minimum LXe subcooling and minimum flow in the purification loop.
\subsubsection{Vacuum-insulated piping and valves}
\label{subsubsec:perf:therm:piping}
In normal operation, the purification loop flow path includes a total length of vacuum-insulated piping of almost \SI{50}{\meter} and about ten cryogenic valves (Sec.~\ref{subsec:loop}). Since the valves are all of the same design, and since the piping is of the same construction and has nearly the same diameter throughout, the thermal performance can be evaluated by estimating the heat input per unit length of piping and per valve.  In principle, this could be performed by measuring the LXe temperature increase along the flow path in a single operating condition. In practice, however, since the temperature increase is small compared to the systematic uncertainty in the temperature measurement arising from both the sensor itself and from its thermal coupling to the piping, that is not possible. This nuisance can be circumvented by using measurements of the increase in temperature between pairs of sensors as a function of LXe flow, as described below.

For a constant heat input per unit length $\dqdlt$, the mean temperature of a fluid increases linearly with distance $x$ along the length of a pipe~\cite{bergman}
\begin{linenomath*}
\begin{equation} \label{eq:dqdl}
	T(x) = T_0 + \dqdl \frac{x}{\rho \dot{V} c} \mathrm{,}
\end{equation}
\end{linenomath*}
where $\rho$ is the fluid density, $\dot{V}$ its volumetric flow, $c$ its specific heat capacity, and where $T_0$ is the temperature at $x = 0$. If each temperature measurement includes an unknown systematic offset, then as the flow becomes large the measured temperature difference should tend toward the difference between offsets since the actual temperature increase becomes negligible. Therefore, it is possible to define the measured temperature change between two sensors along a specific flow path $i$ as
\begin{linenomath*}
\begin{equation} \label{eq:dqdl2}
	\Delta T_i = \Delta T_{\mathrm{sys}, i} + \frac{1}{\rho \dot{V} c} \left({\dqdl L_i + q_{\scriptscriptstyle V} N_{\scriptscriptstyle V,i}}\right),
\end{equation}
\end{linenomath*}
where $\Delta T_{\mathrm{sys}, i}$ is the systematic offset between the two sensors, $L_i$ is the length of pipe, $N_{\scriptscriptstyle V,i}$ is the number of valves, and $q_{\scriptscriptstyle V}$ is the heat input per valve. By choosing flow paths of varying lengths and number of valves, and analyzing the measured temperature increase between each sensor pair over a range of flows, it is possible to simultaneously estimate the offsets, the heat input per unit length, and the heat input per valve.

\begin{figure}[t]
\centering
    \includegraphics[width=\columnwidth]{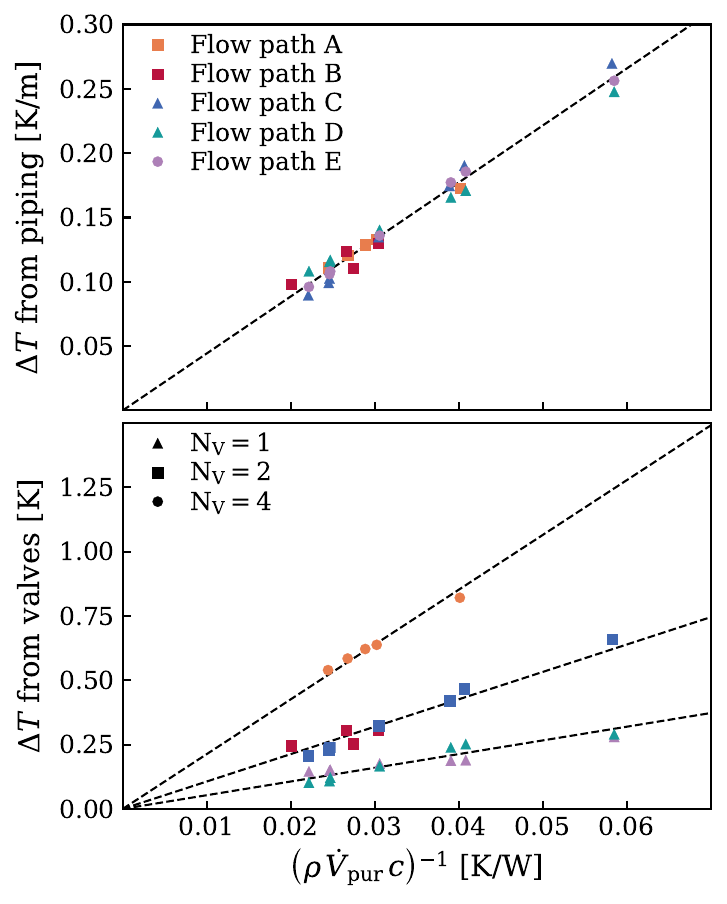}
    \caption{\textbf{Top:} Offset-corrected measured temperature change per unit length attributable to heat input from the piping alone as a function of the inverse heat capacity rate $(\rho \venturiflow c)^{-1}$ for flow path A (orange), B (red), C (blue), D (teal), and E (violet). The dashed line shows the result of the minimization of the set of equations of the form of Eq.~(\ref{eq:dqdl2}) for the five flow paths. Its slope corresponds to $\dqdlt$ and yields an estimate of \SI[per-mode = symbol]{4.4}{\watt\per\meter}. \textbf{Bottom:} Offset-corrected measured temperature change attributable to heat input from valves alone as a function of the inverse heat capacity rate $(\rho \venturiflow c)^{-1}$ for each flow path. The dashed lines show the result of the minimization and have slopes that correspond to $q_{\scriptscriptstyle V}$ times the number of valves $N_\mathrm{V}$ in each flow path. The estimate obtained is \SI{5.3}{\watt} per valve.}
    \label{fig:dqdl}
\end{figure}

The following five flow paths (see Fig.~\ref{fig:pid}) were chosen to perform such a calculation:
\begin{description}[align=left]
  \item [\textit{Flow path A:}] $L_A = \mathrm{\SI{6.6}{\meter}}$,
  \item [\quad TE-272 $\rightarrow$ FV-272, -275, -276, -279 $\rightarrow$ TE-277]
  \item [\textit{Flow path B:}] $L_B = \mathrm{\SI{3.6}{\meter}}$,
  \item [\quad TE-276 $\rightarrow$ FV-278, -284 $\rightarrow$ TE-281]
  \item [\textit{Flow path C:}] $L_C = \mathrm{\SI{3.3}{\meter}}$,
  \item [\quad TE-276 $\rightarrow$ FV-278, -281 $\rightarrow$ TE-274]
  \item [\textit{Flow path D:}] $L_D = \mathrm{\SI{2.7}{\meter}}$,
  \item [\quad TE-274 $\rightarrow$ FV-282 $\rightarrow$ TE-281]
  \item [\textit{Flow path E:}] $L_E = \mathrm{\SI{7.4}{\meter}}$,
   \item [\quad TE-281 $\rightarrow$ FCV-286 $\rightarrow$ TE-282]
\end{description}
These specific flow paths were chosen because they do not have significant overlap and they cover the two sampling modes of the PM (Sec.~\ref{subsec:pm}): flow paths A and B when sampling upstream of the purifier units, and flow paths C and D when sampling downstream. Flow path E is independent of the sampling modes. Figure~\ref{fig:dqdl} shows the result of the calculation. The top plot shows the offset-corrected measured temperature change per unit length attributable to heat input from the piping alone, as a function of $(\rho \venturiflow c)^{-1}$ for each flow path. The dashed line shows the result of the minimization of the set of equations of the form of Eq.~(\ref{eq:dqdl2}) for the five flow paths. Its slope corresponds to $\dqdlt$ and yields an estimate of \SI[per-mode = symbol]{4.4}{\watt\per\meter}. Analogously, the bottom plot shows the offset-corrected measured temperature change attributable to heat input from valves alone, as a function of $(\rho \venturiflow c)^{-1}$ for each flow path. The dashed lines show the result of the minimization and have slopes that correspond to $q_{\scriptscriptstyle V}$ times the number of valves in each flow path. The estimate obtained is \SI{5.3}{\watt} per valve.

The estimate of heat input per unit length of vacuum-insulated piping appears somewhat large compared to the idealized theoretical value of \SI[per-mode = symbol]{\lesssim 1}{\watt \per \meter} that one could expect for highly evacuated MLI~\cite{flynn}. There are several factors that could explain this larger value: the residual pressure in the vacuum insulation since it is only evacuated periodically; the number of field joints between sections of vacuum-insulated piping where the inner pipe is supported by a thin metal sheet in contact with the outer pipe; and the number of penetrations through the MLI to accommodate pressure and temperature sensors, piping tees, etc. It should also be noted that this estimate was performed on relatively more convoluted piping sections than the LXePUR inlet line or the LXe return line~(Fig.~\ref{fig:loop}). The absence of a temperature sensor upstream of the \lntwo-LXe heat exchanger (HX-288) makes the analysis of these piping sections impossible. On the other hand, the heat input per valve appears reasonable when compared to the expectation of \SI{\gtrsim 2}{\watt} solely from solid conduction between the cold valve body and its room temperature flange. In practice, even assuming that the $\dqdlt$ value obtained does not overestimate the heat input for the piping of the entire purification loop, this level of heat input can be compensated by the HX-288 heat exchanger without any issue.
\subsubsection{Purifiers}
\label{subsubsec:perf:therm:purifiers}
The two main components of a purifier unit, the LXe purifier vessel and the particulate filter vessel, have significantly larger diameters than the LXe-carrying piping. This could result in a larger heat input compared to the one from similar piping lengths. The heat input from both vessels of one purifier unit, and the associated piping in between, was estimated using the technique described in the previous section with the single flow path TE-277 $\rightarrow$ F-274 $\rightarrow$ F-275 $\rightarrow$ TE-278. An estimate of \SI{1.7}{\watt} was obtained. No substantial difference was observed in the estimated heat input from Purifier 1 or Purifier 2 (Tab.~\ref{tab:purifier_types}) when either was installed as F-274. This smaller value compared to that obtained for the vacuum-insulated piping is likely due to a combination of factors: the lower pressure in the vacuum insulation enclosure of the purifier unit as it is continuously evacuated with a turbomolecular pumping station; the greater ease at which MLI could be appropriately applied to the two vessels; and the small number of direct heat conduction paths to the room temperature enclosure. 

\begin{figure}[t]
    \centering
    \includegraphics[width=\columnwidth]{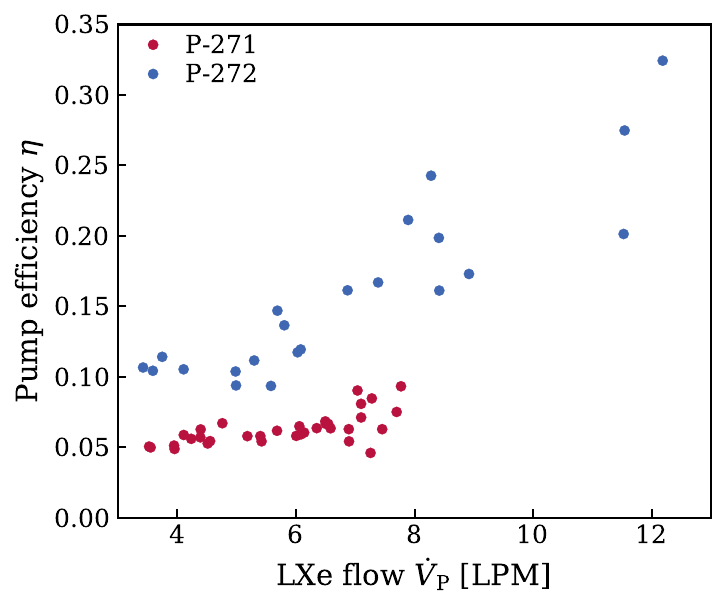}
    \caption{Estimated LXe pump efficiency $\eta$ as a function of flow (\pumpflow) for both P-271 (red) and P-272 (blue) using Eq.~(\ref{eq:pump_efficiency}). The efficiency for P-272 is compatible with the design estimate by the pump manufacturer. The lower efficiency for P-271 is not unexpected given its lower than expected pressure-flow performance (Fig.~\ref{fig:pump_curves}).}
    \label{fig:eta}
\end{figure}
\subsubsection{LXe pump}
\label{subsubsec:perf:therm:pump}
The efficiency $\eta$ of a centrifugal pump is defined as the ratio of useful hydraulic power delivered to the fluid to the input (or shaft) power. Furthermore, the shaft power can be shown to equal the change in specific total enthalpy of the fluid across the pump times the mass flow rate~\cite{pumphandbook}. In the LXePUR system, the hydraulic power delivered to the fluid by the cryogenic liquid pump (P-271/272) is $\pumpflow \pumpdp$ and therefore its efficiency can be calculated from
\begin{linenomath*}
\begin{equation} \label{eq:pump_efficiency}
	\eta = \frac{\pumpflow \pumpdp}{(\rho \pumpflow) \Delta h_{\mathrm{P}}} = \frac{\pumpdp}{\rho \Delta h_{\mathrm{P}}},
\end{equation}
\end{linenomath*}
where $\Delta h_{\mathrm{P}}$ is the specific enthalpy change across the pump. Note that since the pump vessel inlet and outlet pipes are at the same height and have the same diameter (Fig.~\ref{fig:inside_pump}) the change in total enthalpy reduces to change in static enthalpy. Note also that since the pressure delivered by the pump depends on the flow rate (Sec.~\ref{subsubsec:perf:flow:pumps}) then also does the efficiency, explicitly, $\eta = \eta(\pumpflow)$.

The change in specific enthalpy is calculated from the LXe state before and after the pump. For P-271 (P-272), the inlet LXe state is defined by the pressure measured by PT-271, minus the pressure drop across HV-277 (HV-278) (Sec.~\ref{subsubsec:perf:flow:pumps}) and by the temperature measured by TE-288, plus the temperature increase due to heat input from the piping and the valve (Sec.~\ref{subsubsec:perf:therm:piping}).  Similarly, the outlet LXe state for P-271 (P-272) is defined by the pressure measured by PT-275 (PT-276) and by the temperature measured by TE-272 (TE-273), minus the temperature increase due to heat input from both the pump body (conductive) and the piping from the pump to the temperature sensor, corrected by the systematic offset between the two sensors. To determine this systematic offset, an approach was used that avoids the thermal influence of the operational pump. This involves using the cool-down period during the purification loop's start-up. In this period, LXe flows through the sensors while the pump is off, vaporizing as it cools the warm components downstream. At a low flow rate of \SI{\sim 0.1}{\literperminute}, in contrast to the description at the beginning of Sec.~\ref{subsec:perf:flow}, the LXe flowing from the cryostat vessel reaches its saturation temperature rather quickly along the inlet line, as is easily shown by the application of Eq.~(\ref{eq:dqdl}) in these conditions. Since the fluid is at a constant, known temperature as it passes through the pump body and both sensors, the systematic offset is simply calculated as the difference between the two measured temperature values. The data acquired during the LXePUR commissioning periods to determine the characteristic curves of the two LXe pumps (Sec.~\ref{subsubsec:perf:flow:pumps}) is used to calculate the change in specific enthalpy across the pump as a function of flow rate.

The resulting estimated efficiency $\eta$ for both LXe pumps is shown in Fig.~\ref{fig:eta}. The estimated efficiency versus flow for pump P-272 is compatible with the design estimate of $\sim$35\% at \SI{12}{\literperminute}, decreasing to $\sim$25\% at \SI{8}{\literperminute}, from the cryogenic liquid pump manufacturer~\cite{barbernichols}. The lower efficiency estimated for pump P-271 is not unexpected given its pressure-flow performance~(Fig.~\ref{fig:pump_curves}). The differential pressure (\SI{1.7}{\bar}) and pump flow rate (\SI{\sim 5}{\literperminute}) for normal operation mentioned at the end of Sec.~\ref{subsubsec:perf:flow:pumps} lead to heat inputs of \SI{\sim 240}{\watt} and \SI{\sim 120}{\watt} for pumps P-271 and P-272, respectively. Both heat inputs can be compensated by the \lntwo-LXe heat exchanger (HX-288).

\begin{figure}[t]
    \centering
    \includegraphics[width=\columnwidth]{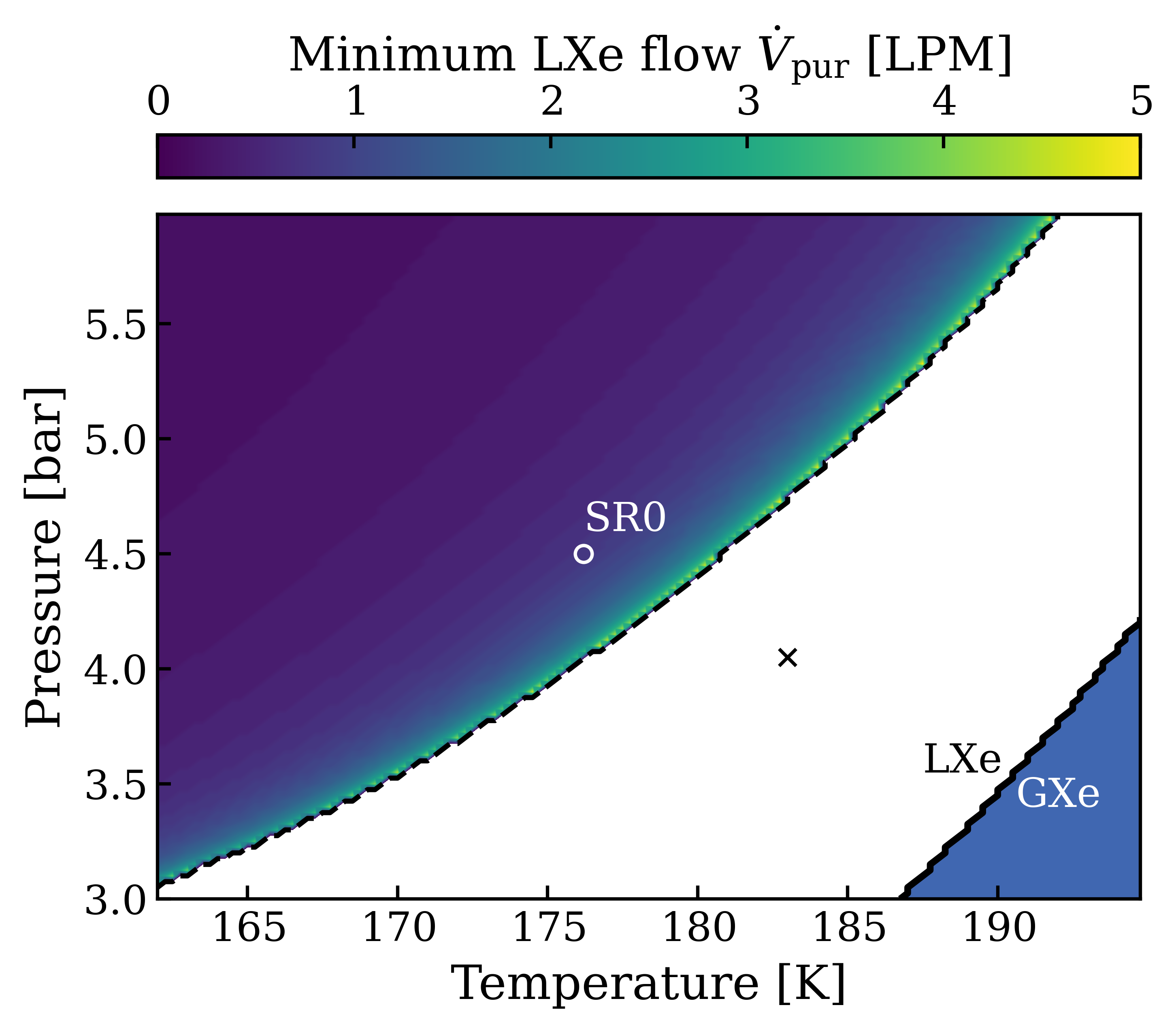}
    \caption{Minimum LXe flow (\venturiflow) for which subcooled LXe reaches the highest point of the LXe return line (Fig.~\ref{fig:loop}) as a function of the fluid state $\left({P, T}\right)$ immediately after the Venturi (FE-288). The more subcooled the LXe is at the outlet of the Venturi, the lower the minimum LXe flow is. On the other hand, for some states (white region) it is not possible to prevent saturation by increasing the flow. This can be compared with the fluid state prevailing during the operation of the LXePUR system at different times: stable operation at \SI{2.0}{\literperminute} during the first science run of XENONnT (SR0, open circle), and a state for which extremely unstable LXe flow was observed during commissioning ($\times$).}
    \label{fig:min_subcooling}
\end{figure}
\subsubsection{Minimum LXe subcooling}
\label{subsubsec:perf:therm:min_subcooling}
As described in Sec.~\ref{subsec:loop} and shown in Fig.~\ref{fig:loop}, after exiting the purifier unit in operation (F-271 and F-272 or F-274 and F-275), the LXe returning to the TPC flows horizontally through the Venturi (FE-288) and then through a combination of vertical and horizontal segments, with a total length of \SI{15.5}{\meter} and a net height of \SI{7.3}{\meter}, before reaching the highest point of the LXe return line. Since the flowing LXe is subject to a small but constant heat input along its path (Sec.~\ref{subsubsec:perf:therm:piping}) and since its pressure decreases substantially as it flows upward (or through a narrowly open valve), it might reach saturation before the highest point of the LXe return line.  Once the flowing LXe reaches saturation, any further reduction of the pressure will lead to the production of GXe via flash evaporation and an evolving two-phase flow pattern~\cite{twophaseflow} as the mixture continues to flow upward. As this can lead to an unsteady LXe mass flow and pressure fluctuations in the cryostat vessel, the LXe exiting the Venturi should be sufficiently subcooled to prevent such a situation. In addition, since the slower the flow the larger the temperature increase will be (Eq.~(\ref{eq:dqdl})), then there is a minimum flow required to prevent saturation. On the other hand, since the pressure drop across a valve is quadratic with the flow (Eq.~(\ref{eq:fv_cv})), there is a limit to how much increasing the flow can help prevent saturation.

This interplay between the minimum subcooling and the minimum flow required can be seen in Fig.~\ref{fig:min_subcooling}. The minimum LXe flow (\venturiflow) for which subcooled LXe reaches the highest point of the LXe return line is calculated as a function of the fluid state $\left({P, T}\right)$ immediately after the Venturi. To perform the calculation, the flow path is discretized into short segments of \SI{1}{cm}. The state of the fluid in a segment is obtained from the state in the upstream segment and any pressure changes due to elevation changes or flow through valves, and the enthalpy increase due to heat influx. The CoolProp package~\cite{coolprop} is used to calculate the temperature in the segment from its enthalpy and pressure, and to determine whether saturation has been reached. If it has, then the calculation is repeated at a higher LXe flow, up to a maximum of \SI{5}{\literperminute}. Here the flow control valve FCV-286 is assumed to be fully opened.

As expected, the minimum LXe flow decreases with increasing subcooling, or in other words, as the operating state moves further away from the saturation curve. The states in the white region in Fig.~\ref{fig:min_subcooling} are not subcooled enough and lead to LXe saturation before the highest point of the LXe return line. This calculation can be compared with operating conditions of the LXePUR system using the temperature sensor TE-281, installed immediately after the Venturi (FE-288), and the pressure transducer PT-286, after adjusting its value for the pressure drop across the Venturi tube (Fig.~\ref{fig:venturi_dp}). The open circle in Fig.~\ref{fig:min_subcooling} shows the fluid state near TE-281 during the first science run of XENONnT~\cite{PhysRevLett.129.161805} during which the system operated stably at \SI{2.0}{\literperminute}. A state for which extremely unstable LXe flow was observed during commissioning is also shown ($\times$). In practice, the flow control valve FCV-286 is partially closed to divert a fraction of the purified LXe flow to the radon removal system (Sec.~\ref{subsec:loop} and ~\ref{subsec:perf:flow}). As discussed above, the effect would be to extend the white region further to the left due to the increased pressure drop across the valve. 

Note that not all fluid states in Fig.~\ref{fig:min_subcooling} can be obtained immediately after the Venturi with the LXePUR system. The minimum set temperature (TIC-288) of the \lntwo-LXe heat exchanger (HX-288) for stable operation is about \SI{7}{\kelvin} above the triple point (\SI{161.4}{\kelvin}) and the LXe pump and piping will add heat to the fluid (Sec.~\ref{subsubsec:perf:therm:pump}). Also, the LXe pump inlet pressure at nominal cryostat vessel pressure being \SI{\sim 3.4}{\bar} and the maximum deliverable differential pressure of \SI{\sim 1.9}{\bar} (Sec.~\ref{subsubsec:perf:flow:pumps}) for P-271 make it difficult to have pressures greater than \SI{\sim 5}{\bar} immediately after the Venturi. Similarly, since the LXe pressure at the highest point of the LXe return line should be greater than the cryostat vessel pressure, then a pressure \SI{\gtrsim 4}{\bar} is required immediately after the Venturi. Given all these constraints, the minimum stable LXe flow is about \SI{0.9}{\literperminute}.
\subsection{Purification performance}
\label{subsec:perf:pur}
The purification performance of the LXePUR system is manifestly of prime interest. The main criteria that are useful to assess the performance of a purification system are: the time required to achieve a target purity, either after an initial filling of the volume to be purified or after a maintenance interruption; the lowest impurity concentration attainable in normal operation; and the maximum time interval between maintenance operations. As can be seen from Eqs.~\ref{eq:purity_model} and \ref{eq:equilibium_purity}, the first two criteria depend directly on the volume exchange time $\tau$ and the impurity removal efficiency $\epsilon$ of the system. In addition, the first depends on the initial purity level before the purification system is started while the second depends on the rate at which impurities are entering the volume to be purified, $\Lambda$, both of which are properties of the volume to be purified. The third criterion will depend both on the total impurity sorption capacity of the purifier used and on the other components of the purification system, such as LXe pump maintenance.

In the specific case of the XENONnT experiment, the LXe volume is purified both by the LXePUR system and the GXePUR system (Sec.~\ref{sec:intro}), albeit with a much longer time constant with the latter, and Eq.~(\ref{eq:purity_model}) becomes
\begin{linenomath*}
\begin{equation} \label{eq:purity_model_with_gpur}
	n \frac{dx}{dt} = \Lambda - \frac{\epsilon n}{\tau}x - \frac{\epsilon' n}{\tau'}x,
\end{equation}
\end{linenomath*}
where $\epsilon$ and $\tau$ are the \otwo removal efficiency of the purifier and the volume exchange time associated with the LXePUR system, respectively, whereas the primed quantities are for the GXePUR system. The LXe volume purified has a total mass $m$ of $\sim$\nTtotalLXeMass, \nTCryostatLXeMass of which are contained in the cryostat, \SI{\sim 50}{\kilo\gram} in the LXePUR purification loop, and \SI{\sim 60}{\kilo\gram} in the radon removal system. In what follows, the \otwo removal efficiency of the heated getter gas purifiers of the GXePUR system is assumed to be $\epsilon' = 1$~\cite{dobi}. In contrast, the \otwo removal efficiency of a given purifier of the LXePUR system, a priori unknown, is to be inferred. This can be achieved in two different ways: indirectly, via Eq.~(\ref{eq:purity_model_with_gpur}), by modeling the time dependence of the \otwo mole fraction $x$ calculated from the PM measurements of the electron lifetime using Eq.~(\ref{eq:elconversion}); or more directly, by using the LXePUR system capability to measure the LXe purity either upstream or downstream of its purifiers (Sec.~\ref{subsec:pm}). Specifically, assuming that any \otwo sources between the PM and the purifier (and vice versa) are negligible implies that the \otwo mole fraction at the purifier inlet (outlet) is the same as the mole fraction in the PM when it is sampling the LXe upstream (downstream) of the purifier.  Consequently, in Eq.~(\ref{eq:purifier_eff}), $x_{\mathrm{pur}}(z=0) \propto \tau_{e,\mathrm{u}}^{-1}$ and $x_{\mathrm{pur}}(z=L) \propto \tau_{e,\mathrm{d}}^{-1}$, where $\tau_{e,\mathrm{u}}$ and $\tau_{e,\mathrm{d}}$ correspond to PM electron lifetime measurements when sampling in the upstream and downstream configuration, respectively, and hence
\begin{linenomath*}
\begin{equation} \label{eq:eff_pre_post}
	\epsilon = 1 - \frac{\tau_{e,\mathrm{u}}}{\tau_{e,\mathrm{d}}}.
\end{equation}
\end{linenomath*}
Although not explicitly mentioned above, deriving the \otwo mole fraction in the detector $x$ from PM electron lifetime measurements upstream of the purifier unit also relies on the assumption that \otwo sources between the cryostat vessel and the PM are negligible.

Note that in the indirect approach (Eq.~(\ref{eq:purity_model_with_gpur})), $x$ has to vary appreciably with time for the method to provide an accurate estimate of $\epsilon$. Additionally, any time dependence in $\Lambda$, that is unaccounted for, can lead to a systematic error on $\epsilon$. In the direct approach (Eq.~(\ref{eq:eff_pre_post})), the purity in the detector and at the outlet of the purifier unit should both be within the range that the PM can measure precisely in order to obtain a value for $\epsilon$ with a small systematic uncertainty. The validity of the assumption of negligible \otwo sources upstream and downstream of the purifier could also lead to systematic errors.

The \otwo removal efficiency and total sorption capacity of the prototype purifiers (C1-3, Sec.~\ref{subsubsec:perf:flow:purifiers}) tested with the LXePUR system during its two commissioning periods are discussed in Sec.~\ref{subsubsection:perf:pur:prototypes} while those of the two purifier types in use, Purifiers 1 and 2 (Tab.~\ref{tab:purifier_types}), are discussed in Secs.~\ref{subsubsection:perf:pur:q5} and \ref{subsubsection:perf:pur:st707}, respectively.
\subsubsection{Prototype purifiers}
\label{subsubsection:perf:pur:prototypes}
The purification performance of the prototype purifiers C1-3 described in Sec.~\ref{subsubsec:perf:flow:purifiers} was evaluated during the two commissioning test sequences of the LXePUR system. An initial hypothesis was that it might be possible to achieve a relatively high \otwo removal efficiency with an acceptable radiopurity level by using a small mass (tens of grams) of the Q5 sorbent. However, the purification tests performed with prototype purifiers C1 and C2 revealed that this approach was not viable: at a LXe flow rate of \SI{\sim 2}{\literperminute} the \otwo removal efficiencies obtained were low, \SI{\sim 10}{\percent} and \SI{\sim 30}{\percent}, respectively. Furthermore, a strong decrease of the efficiency after a short service time was observed, on the order of a factor two within the course of a day, in both cases after having captured \SI{< 1}{\milli\gram} of \otwo. This performance was difficult to reconcile with prior results~\cite{xeclipse} and appeared more likely to have been due to a significant fraction of the small mass of sorbent being passivated with \water in this initial purification phase. The purification performance of prototype purifier C3 (St707) showed a higher initial \otwo removal efficiency of \SI{\sim 70}{\percent} at \SI{1.5}{\literperminute}. Note that the sorbent mass in prototype purifier C3 was considerably larger than in C1 and C2 (Tab.~\ref{tab:purifier_types_proto}). It nevertheless exhibited a similar behavior of strong decrease of efficiency over a short service period but at a slightly slower rate. Thus, it seemed that achieving an \otwo removal efficiency of \SI{\sim 100}{\percent} with a large enough sorption capacity for the initial purification phase would require a relatively large mass of St707 sorbent ($>\SI{5}{kg}$).

Throughout these prototype purifier tests, an interesting effect was observed. In each case, immediately after the purifier was put in service, the initial \otwo removal efficiency was evaluated via Eq.~(\ref{eq:eff_pre_post}) as a function of LXe flow rate. The LXe flow rate was increased in steps up to \SI{\sim 2}{\literperminute} and then decreased in steps back to the initial flow rate, with $\tau_{e,\mathrm{d}}$ measured at each step for \SI{\sim 1}{\hour}. The expectation was that the outlet concentration of the purifier would worsen with increasing LXe flow rates due to a decrease in contact time in the purifier. While this was indeed observed, the $\tau_{e,\mathrm{d}}$ value measured when returning to a previous (lower) flow rate was always worse than before. The most reasonable explanation for such an observation seems to be that circulating LXe with a significantly lower purity can contaminate the internal surfaces of the piping between the purifier and the PM. When cleaner LXe is circulated, these surfaces will shed impurities and it will take time until the actual purifier outlet concentration is measured at the PM.

The performance results obtained with these prototype purifiers lead to the dual purifiers design choice (Sec.~\ref{subsec:purifiers}): a high-efficiency purifier with a large sorption capacity intended for use during the initial purification phase, and for which the radiopurity is not critical, realizable with a Q5 sorbent bed; and a purifier with a more modest efficiency and sorption capacity but a low enough \rnbkg activity, for use during normal operation, realizable with an St707 sorbent bed.

\subsubsection{Purifier 1 performance}
\label{subsubsection:perf:pur:q5}
\begin{figure}[!h]
    \centering
    \includegraphics[width=\columnwidth]{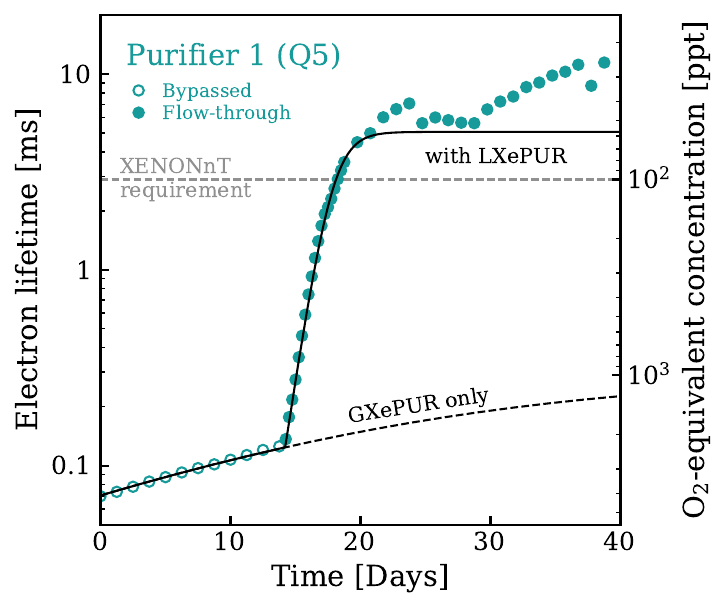}
    \caption{Electron lifetime improvement before (open circles) and after (closed circles) initiating purification with Purifier 1 (Q5). The LXe purity improves slowly while purifying with the GXePUR system only but does so drastically in comparison once Purifier 1 of the LXePUR system is put in service ($t = \SI{14.2}{\day}$). The observed improvement is well described (solid line) by Eq.~\ref{eq:purity_model_with_gpur} assuming a constant rate of release of \otwo into the LXe volume and a purifier efficiency of 100\%. The XENONnT purity goal would have remained out of reach without using the LXePUR system (dashed line).}
    \label{fig:q5_evol}
\end{figure}
The most suitable period to assess the purification performance of Purifier 1 (Q5, Tab.~\ref{tab:purifier_types}) is during the commissioning period of XENONnT, shortly after the cryostat inner vessel and the TPC were filled with LXe. During the filling, the increasing LXe volume accumulates impurities, and the \otwo concentration is at its highest at the end of the operation before purification is initiated. A few days after the XENONnT cryostat filling was completed, the LXe pump P-271 was started using a valve configuration such that LXe was flowing through the PM but bypassing the purifier units. The initial electron lifetime measured was $\SI{\sim 71}{\micro\second}$ ($x(t=0) = \SI{4}{\ppb}$). Shortly thereafter, purification of the LXe volume was initiated with the GXePUR system (Sec.~\ref{sec:intro}) at a mass flow rate $\dot{m}' = \SI{58}{\standardliterperminute}$, corresponding to a volume exchange time of $\tau' = m/\dot{m}' = \SI{17.8}{\day}$. After \SI{14.2}{\day} of purification with the GXePUR system only the LXePUR system valve configuration was changed so that LXe would flow through Purifier 1 at a rate of $\venturiflow = \SI{2}{\literperminute}$, corresponding to a volume exchange time of $\tau = m/(\rho \venturiflow) = \SI{1}{\day}$. Note that throughout, except for a short test period (see below), the GXe volume of the cryostat was also purified in parallel with the GXePUR system at a small flow rate of \SI{2}{\standardliterperminute}. This corresponds to a short volume exchange time (\SI{\sim 29}{\hour}) given the considerably smaller quantity of xenon contained in this volume (\SI{\sim 20}{\kilo\gram}).

Figure~\ref{fig:q5_evol} shows the improvement in the measured electron lifetime during the purification period with the GXePUR system only (open circles) and the substantial effect of initiating purification with the LXePUR system (closed circles) using Purifier 1 (Q5). After two weeks of purification with the GXePUR system the electron lifetime had increased by \SI{\sim 40}{\percent} to \SI{\sim 100}{\micro\second}. With the LXePUR system and Purifier 1, it further improved by a factor of $\sim 60$ in one week. A measurement of the electron lifetime downstream of Purifier 1 was performed a few hours after diverting the LXe flow to the purifier and yielded an \otwo removal efficiency of $\epsilon \sim 100\%$ via Eq.~(\ref{eq:eff_pre_post}). A more robust assessment of the \otwo removal efficiency of Purifier 1 can be obtained by exploring numerical solutions of Eq.~(\ref{eq:purity_model_with_gpur}).  The behavior observed in the first three weeks is well described by a solution (solid line in Fig.~\ref{fig:q5_evol}) that assumes, in addition to the other considerations discussed above, a constant \otwo source of $\Lambda = \SI[per-mode=symbol]{0.12}{\milli\gram\per\day}$ and a constant purifier efficiency of 100\%.

In the period that follows, the measured electron lifetime is consistently higher than this prediction and continues to increase, which is most likely due to a gradual decrease of $\Lambda$ with time. This is expected since the \otwo sources deplete over time, leading to a decrease in the rate of release of impurities~\footnote{For a discussion of the time dependence of different
mechanisms of gas release from solids, see e.g.~\cite{ohanlon} ch.~4}. Between days 25 and 30, a test was performed where the purification of the GXe volume of the cryostat was temporarily interrupted. The slight decrease and flattening observed is a strong indication that a significant contribution to $\Lambda$ was from \otwo sources located in the GXe volume. As a comparison, Fig.~\ref{fig:q5_evol} also shows a solution of Eq.~(\ref{eq:purity_model_with_gpur}) with GXePUR only (dashed line). In this case, the time to reach equilibrium purity (Eq.~(\ref{eq:equilibium_purity})) is considerably longer, and its value is worse by a factor of $(\tau'+\tau)/\tau$.

As discussed in Sec.~\ref{subsec:purifiers}, the mass of Q5 sorbent in Purifier 1 was chosen such that its expected \otwo sorption capacity would be much greater than \SI{100}{\milli\gram}. Given that only \SI{\sim 4}{\milli\gram} of \otwo were captured by Purifier 1 during this initial purification phase, and given that the purifier was in service for a relatively short period afterwards, no additional information on its sorption capacity can be derived.

\begin{figure}
    \centering
    \includegraphics[width=\columnwidth]{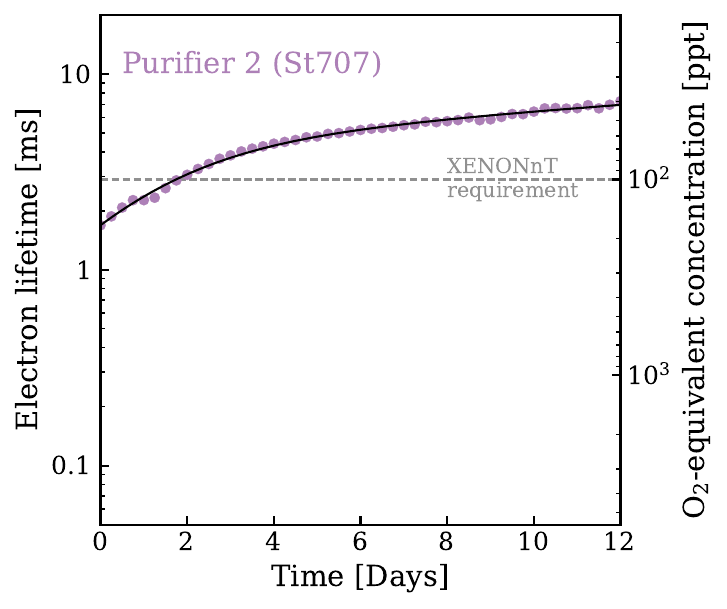}
    \caption{Electron lifetime improvement after Purifier 2 (St707) was put in service (closed circles), following the initial purification phase with Purifier 1 (Fig.~\ref{fig:q5_evol}) and a brief (\SI{\sim 10}{\day}) interruption of the LXePUR system (not shown). The observed improvement is described (solid line) by Eq.~(\ref{eq:purity_model_with_gpur}) assuming a slowly decreasing rate of release of \otwo into the LXe volume (Eq.~(\ref{eq:lambda_time})) and a constant purifier efficiency of $\epsilon=\negefficiencyfit$.}
    \label{fig:st707_evol}
\end{figure}
\subsubsection{Purifier 2 performance}
\label{subsubsection:perf:pur:st707}
The most suitable period to assess the purification performance of Purifier 2 (St707, Tab.~\ref{tab:purifier_types}) in terms of its initial \otwo removal efficiency is also immediately after it is put in service, when the LXe purity varies substantially. This was performed a few weeks after the initial purification phase shown in Fig.~\ref{fig:q5_evol}. The improvement in the electron lifetime after Purifier 2 was put in service, as measured by the PM, is presented in Fig.~\ref{fig:st707_evol}. Note that prior to the use of Purifier 2 shown here, purification with the LXePUR system was interrupted for a period of about ten days, which accounts for the lower initial electron lifetime in Fig.~\ref{fig:st707_evol}. The LXe volume purification flow rates used were slightly different than those of Sec.~\ref{subsubsection:perf:pur:q5}, $\dot{m}' = \SI{61}{\standardliterperminute}$ and $\venturiflow = \SI{2.1}{\literperminute}$, while the cryostat vessel GXe volume purification mass flow was increased to \SI{17}{\standardliterperminute}. The electron lifetime improved from \SI{\sim 1.7}{\milli\second} (\SI{\sim 170}{\ppt}) to \SI{\sim 7}{\milli\second} (\SI{\sim 41}{\ppt}) in \SI{12}{\day}. As was observed in Fig.~\ref{fig:q5_evol}, the rate of release of \otwo into the LXe volume appears to decrease in time, leading to a slow improvement of the purity instead of a plateau. The \otwo removal efficiency of Purifier 2 can be determined via the indirect method. The time dependence of the \otwo source is assumed to take the form
\begin{linenomath*}
\begin{equation} \label{eq:lambda_time}
	\Lambda (t) = \frac{\Lambda (0)}{1 + t/T_{1/2}},
\end{equation}
\end{linenomath*}
the same as was used in \cite{xeclipse}, where $\Lambda (0)$ is the initial rate of release, and $T_{1/2}$ corresponds to the time after which the rate will have reduced to half its initial value. Again, the purifier efficiency $\epsilon$ is assumed to remain constant over the period considered here. 
Figure~\ref{fig:st707_evol} (solid line) shows a solution of Eq.~(\ref{eq:purity_model_with_gpur}) that describes the measured behavior, where $\Lambda (0) = \SI[per-mode=symbol]{86}{\micro\gram\per\day}$, $T_{1/2} = \SI{17}{\day}$, and $\epsilon = \SI{55}{\percent}$.

A measurement of the \otwo removal efficiency of Purifier 2 via the direct method (Eq.~(\ref{eq:eff_pre_post})) was also attempted shortly after it was first put in service. However, the effect described in Sec.~\ref{subsubsection:perf:pur:prototypes}, where PM electron lifetime values are systematically affected for some time after changing from the upstream to the downstream sampling configuration, hindered the measurement. A second attempt was made after about four months of service of Purifier 2 where a longer sampling period (\SI{\sim 4}{\hour}) was used. The value obtained was $\epsilon=0.56^{+0.08}_{-0.06}$, compatible with the efficiency inferred from the time evolution model in Fig.~\ref{fig:st707_evol}. The uncertainty here comes from the uncertainty in the $\tau_{e,\mathrm{d}}$ measurement, which is appreciable when the electron lifetime is above \SI{10}{\milli\second}, as was the case here.

The flow dependence of the \otwo removal efficiency of Purifier 2 was investigated by operating the purification loop at two different flow rates, $\SI{2.1}{\literperminute}$ and $\SI{1.5}{\literperminute}$.  The change was performed in a period where $\Lambda$ was constant, inferred from the constancy of the electron lifetime. No measurable change in the LXe equilibrium purity (Eq.~(\ref{eq:equilibium_purity})) was observed with the change of flow rate, and thus, at this flow regime ($\mathcal{O}(\SI{1}{\literperminute})$), the \otwo removal efficiency of the purifier appears to vary inversely with the flow rate. The St707 sorbent mass required in order to have a purifier with an \otwo removal efficiency of 100\% at $\SI{\sim 2}{\literperminute}$ would therefore be at least $1/\epsilon$ ($\approx$1.8) times that of Purifier 2 (Tab.~\ref{tab:purifier_types}), that is, \SI{> 5.5}{\kilo\gram}.

\begin{figure*}[!h]
    \centering
    \includegraphics[width=\linewidth]{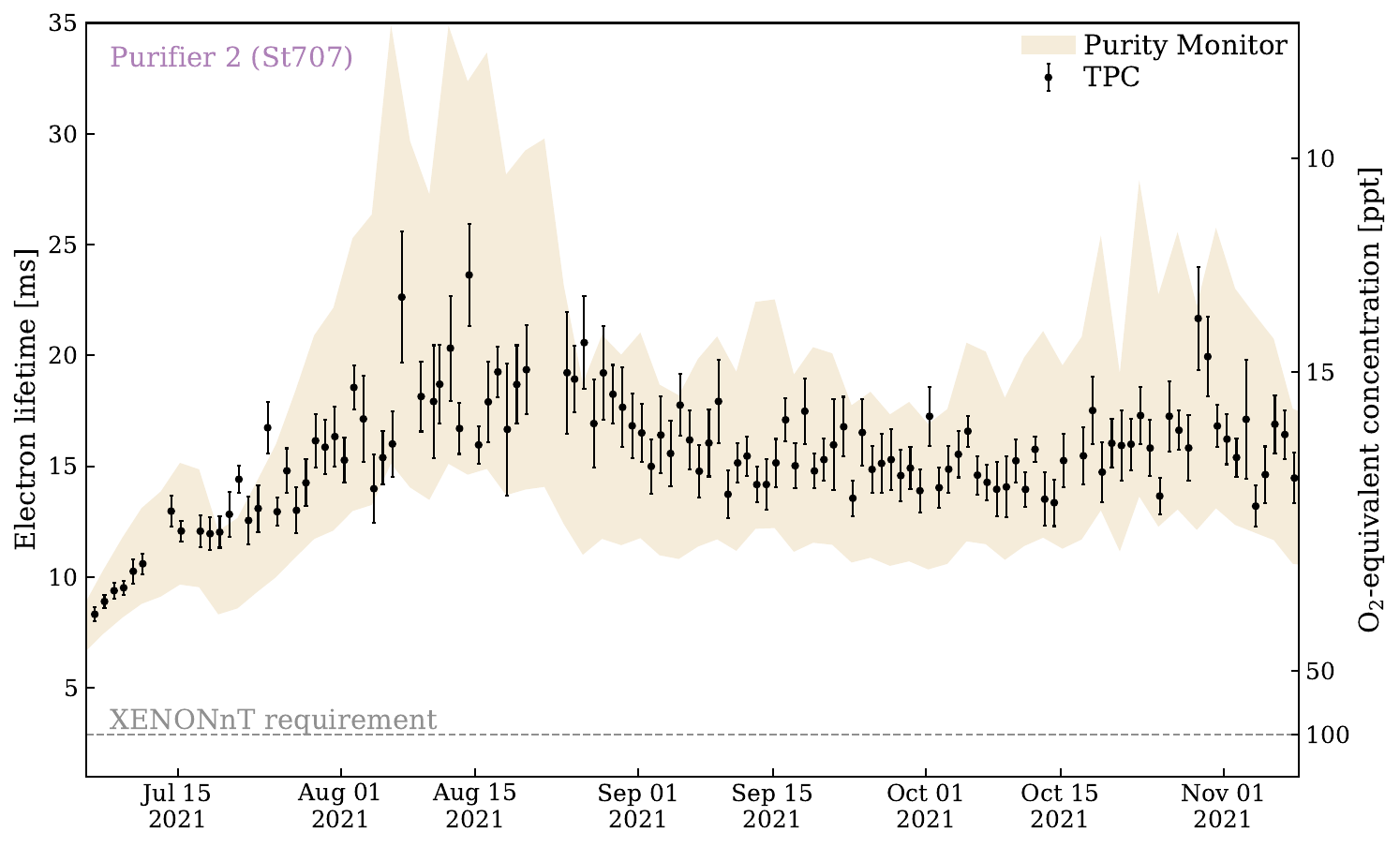}
    \caption{Electron lifetime measurements from both the PM (beige band) and the TPC (black dots) over several months of operation of the LXePUR system with Purifier 2 (St707). The band contains 68\% of the PM measurements. The electron lifetime in the TPC is calculated from \rnbkg alpha decays~\cite{analysispaper1,XENONnT:2023dvq}. An electron lifetime in excess of \SI{15}{\milli\second} was achieved for most of the period. No clear sign of decrease of the \otwo removal efficiency of the purifier was observed leading to a lower bound of~\negamtcaptured for its sorption capacity.}
    \label{fig:sr0_lifetime}
\end{figure*}

The \otwo sorption capacity of Purifier 2, or its \otwo removal efficiency dependence on the sorbed \otwo quantity, can only be probed with long periods of continuous operation of the LXePUR system.  Figure~\ref{fig:sr0_lifetime} shows the long-term variation of the LXe purity in XENONnT, inferred from the PM (beige band) and TPC (black dots) measurements of the electron lifetime, over a period that coincides with the first science run of the experiment~\cite{PhysRevLett.129.161805,xenonnt:sr0wimp} and where Purifier 2 was in use continuously. The beige band contains 68\% of PM electron lifetime measurements. Most of the PM electron lifetime measurements here are larger than the higher limit (\SI{10}{\milli\second}) for precise measurements (\SI{\leq 20}{\percent} relative uncertainty) discussed in Sec.~\ref{subsec:pm}, hence the larger variation in measurements. It is worth noting that given the functional dependence of the PM electron lifetime on the charge signal ratio (Eq.~(\ref{eq:pm_elife})), its relative uncertainty diverges once the LXe purity becomes high and the signal ratio approaches unity. The electron lifetime in the TPC is calculated from \rnbkg alpha decays, which are distributed uniformly throughout the TPC volume~\cite{analysispaper1,XENONnT:2023dvq}. Note that the drift field of the TPC (\SI{\sim 23}{\volt\per\centi\meter}) is not exactly the same as that of the PM (\SI{100}{\volt\per\centi\meter}). The two electron lifetime measurement methods were nevertheless found to yield results consistent with each other. Assuming the \otwo removal efficiency of $\epsilon=\negefficiencyfit$ measured above, and including purifier service time prior to the period shown in Fig.~\ref{fig:sr0_lifetime}, leads to an estimated total \otwo mass captured by Purifier 2 of~\negamtcaptured. Since no clear decrease in efficiency is observed, this constitutes a lower bound for the sorption capacity of this purifier. 
\section{Conclusion}
\label{sec:conclusion}
XENONnT relies on a novel purification system (LXePUR) to maintain the purity of its liquid xenon. Unlike traditional methods that work with gas at high temperature, LXePUR operates at a cryogenic temperature around \SI{175}{K}, directly processing liquid xenon, which is about 300 times denser than gaseous xenon to remove impurities such as oxygen (\otwo). These impurities can significantly diminish the charge signals detected in XENONnT's dual-phase time projection chamber (TPC) by electron attachment. This approach enables rapid purification of the \nTtotalLXeMass xenon inventory, with a purification time constant of \SI{\sim 25}{\hour} at a volumetric flow rate of \SI{\sim 2}{\literperminute}.  The LXePUR continuously circulates the xenon: a cryogenic pump drives it through a purity monitor, a purifier vessel, and a particulate filter before returning it to the TPC. 

The cryogenic liquid pumps feature a magnetically-coupled shaft that completely isolates the motor and its windings from the rotating shaft and the impeller, and therefore from the fluid being pumped, minimizing the impurities introduced into the system. Both are capable of achieving the necessary pressure differential of up to \SI{2}{bar} at flow rates between 4 and 6 \si{\literperminute}, with about half the flow bypassing the purification loop and being fed back at the pump inlet. A heat exchanger, using liquid nitrogen as a cold fluid, subcools the xenon entering the pump and prevents the bypass flow from continuously warming up. At the normal-operation pump flow of about 4–6 \si{\literperminute}, P-272 demonstrates an efficiency ($\eta$) of approximately \SI{10}{\percent}, aligning closely with manufacturer design estimates, while P-271 is approximately \SI{50}{\percent} lower. However, even with this reduced efficiency, the heat exchanger is capable of maintaining the heat input from the LXePUR system into the cryostat housing the TPC as a net negative.

The purity monitor (PM), based on a double-grid ionization chamber design, continuously monitors and directly measures the achieved liquid xenon purity in terms of the electron lifetime $\tau_e$. Due to variations in attachment rate constants $k$ between different impurities, XENONnT prioritizes monitoring \otwo, the dominant impurity within the detector. To ensure optimal operation of the TPC, the XENONnT baseline design required an \otwo-equivalent concentration below \SI{100}{ppt}. The measured electron lifetime allows for the calculation of the equivalent concentration of \otwo present. A precise measurement (\SI{\leq 20}{\percent} relative uncertainty) of \otwo-equivalent concentrations down to \SI{29}{\ppt}, corresponding to an electron lifetime of \SI{10}{\milli\second}, was achieved with the PM. 

The purification strategy employs a two-step approach: For an initial cleanup, a purifier with a copper catalyst on high-surface-area alumina support (Purifier 1, Q5) removes impurities from the liquid xenon bulk with nearly \SI{100}{\percent} efficiency, ideal for initial detector filling or after long-term maintenance. During XENONnT's commissioning the electron lifetime increased from \SI{\sim 100}{\micro\second} to \SI{\sim6 }{\milli\second} within one week. However, the significant \rnbkg emanation rate of \SI{180}{\milli \becquerel} from Purifier 1 leads to increased background in the detector. In a second step, a purifier with sintered pellets of non-evaporable getter alloy (Purifier 2, St707) ensures sustained purity with negligible background impact due to its \rnbkg emanation rate of only \SI{0.24}{\milli \becquerel}. Purifier 2 exhibits a \SI{55}{\percent} efficiency, sufficient for normal operation: XENONnT achieved an \otwo-equivalent concentration in liquid xenon below \SI{19}{\ppt} during its first science run, corresponding to an electron lifetime of about \SI{15}{\milli\second}. This exceptional purity allows \SI{86}{\percent} of generated electrons to survive the \SI{2.2}{\milli\second} full drift across the TPC at a drift field of \SI{\sim 23}{\volt\per\centi\meter}, enhancing the experiment’s dark matter sensitivity. Purifier 2 exhibited no signs of degradation during the science run. This leads to an estimated total \otwo mass captured of \negamtcaptured, a conservative value for its sorption capacity.

The pressure-flow performance of the cryogenic liquid pumps and the flow resistance of the main components of the system, as well as their thermal performance, were quantitatively evaluated and modeled during the commissioning of the system. The resulting models make it possible to predict the performance in conditions outside of those tested and to explore the capabilities of such a system beyond the requirements of XENONnT. The maximum achievable purification mass flow rate is not limited by the pump performance and could be increased by reducing the flow resistance of the purifiers and the Venturi flow meter. This could be accomplished by installing purifiers with a larger cross-section and volume, and by replacing the Venturi tube with a Coriolis flow meter. The \otwo removal efficiency of the St707-based purifier could be increased by increasing the liquid xenon contact time or using smaller diameter pellets. Therefore, the technology developed in this work can be scaled up for future experiments with liquid xenon such as the XLZD detector featuring xenon targets of up to 100 tonnes~\cite{XLZD:2024gxx}.
\begin{acknowledgements}
We gratefully acknowledge support from the National Science Foundation, Swiss National Science Foundation, German Ministry for Education and Research, Max Planck Gesellschaft, Deutsche Forschungsgemeinschaft, Helmholtz Association, Dutch Research Council (NWO), Funda\c{c}\~ao para a Ci\^encia e Tecnologia, Weizmann Institute of Science, Binational Science Foundation, R\'egion des Pays de la Loire, Knut and Alice Wallenberg Foundation, Kavli Foundation, JSPS Kakenhi, JST FOREST Program, and ERAN in Japan, Tsinghua University Initiative Scientific Research Program, National Natural Science Foundation of China, Ministry of Education of China, DIM-ACAV+ R\'egion Ile-de-France, and Istituto Nazionale di Fisica Nucleare. This project has received funding/support from the European Union's Horizon 2020 and Horizon Europe research and innovation programs under the Marie Sk\l{}odowska-Curie grant agreements No 860881-HIDDeN and No 101081465-AUFRANDE.

We gratefully acknowledge support for providing computing and data-processing resources of the Open Science Pool and the European Grid Initiative, at the following computing centers: the CNRS/IN2P3 (Lyon - France), the Dutch national e-infrastructure with the support of SURF Cooperative, the Nikhef Data-Processing Facility (Amsterdam - Netherlands), the INFN-CNAF (Bologna - Italy), the San Diego Supercomputer Center (San Diego - USA) and the Enrico Fermi Institute (Chicago - USA). We acknowledge the support of the Research Computing Center (RCC) at The University of Chicago for providing computing resources for data analysis.

We thank the INFN Laboratori Nazionali del Gran Sasso for hosting and supporting the XENON project.
\end{acknowledgements}
\bibliographystyle{unsrturl}
\bibliography{lxepur} 
\end{document}

%% file: Sep26-LXePur.tex





\author{E.~Aprile\orcidlink{0000-0001-6595-7098}\thanksref{addr3}
\and
J.~Aalbers\orcidlink{0000-0003-0030-0030}\thanksref{addr28}
\and
K.~Abe\orcidlink{0009-0000-9620-788X}\thanksref{addr23}
\and
M.~Abu~Rmilah\orcidlink{0009-0007-9750-6655}\thanksref{addr16}
\and
M.~Adrover\thanksref{addr17}
\and
S.~Ahmed~Maouloud\orcidlink{0000-0002-0844-4576}\thanksref{addr18}
\and
L.~Althueser\orcidlink{0000-0002-5468-4298}\thanksref{addr7}
\and
B.~Andrieu\orcidlink{0009-0002-6485-4163}\thanksref{addr18}
\and
E.~Angelino\orcidlink{0000-0002-6695-4355}\thanksref{addr4,addr1}
\and
D.~Ant\'on~Martin\orcidlink{0000-0001-7725-5552}\thanksref{addr1}
% ...
(XENON Collaboration\thanksref{xenonemail}). }
\newcommand{\bologna}{Department of Physics and Astronomy, University of Bologna and INFN-Bologna, 40126 Bologna, Italy}
\newcommand{\chicago}{Department of Physics, Enrico Fermi Institute \& Kavli Institute for Cosmological Physics, University of Chicago, Chicago, IL 60637, USA}
\newcommand{\coimbra}{LIBPhys, Department of Physics, University of Coimbra, 3004-516 Coimbra, Portugal}
\newcommand{\columbia}{Physics Department, Columbia University, New York, NY 10027, USA}
\newcommand{\lngs}{INFN-Laboratori Nazionali del Gran Sasso and Gran Sasso Science Institute, 67100 L'Aquila, Italy}
\newcommand{\mainz}{Institut f\"ur Physik \& Exzellenzcluster PRISMA$^{+}$, Johannes Gutenberg-Universit\"at Mainz, 55099 Mainz, Germany}
\newcommand{\mpik}{Max-Planck-Institut f\"ur Kernphysik, 69117 Heidelberg, Germany}
\newcommand{\munster}{Institut f\"ur Kernphysik, University of M\"unster, 48149 M\"unster, Germany}
\newcommand{\nikhef}{Nikhef and the University of Amsterdam, Science Park, 1098XG Amsterdam, Netherlands}
\newcommand{\nyuad}{New York University Abu Dhabi - Center for Astro, Particle and Planetary Physics, Abu Dhabi, United Arab Emirates}
\newcommand{\purdue}{Department of Physics and Astronomy, Purdue University, West Lafayette, IN 47907, USA}
\newcommand{\rice}{Department of Physics and Astronomy, Rice University, Houston, TX 77005, USA}
\newcommand{\stockholm}{Oskar Klein Centre, Department of Physics, Stockholm University, AlbaNova, Stockholm SE-10691, Sweden}
\newcommand{\subatech}{SUBATECH, IMT Atlantique, CNRS/IN2P3, Nantes Universit\'e, Nantes 44307, France}
\newcommand{\torino}{INAF-Astrophysical Observatory of Torino, Department of Physics, University  of  Torino and  INFN-Torino,  10125  Torino,  Italy}
\newcommand{\ucsd}{Department of Physics, University of California San Diego, La Jolla, CA 92093, USA}
\newcommand{\wis}{Department of Particle Physics and Astrophysics, Weizmann Institute of Science, Rehovot 7610001, Israel}
\newcommand{\zurich}{Physik-Institut, University of Z\"urich, 8057  Z\"urich, Switzerland}
\newcommand{\paris}{LPNHE, Sorbonne Universit\'{e}, CNRS/IN2P3, 75005 Paris, France}
\newcommand{\freiburg}{Physikalisches Institut, Universit\"at Freiburg, 79104 Freiburg, Germany}
\newcommand{\napels}{Department of Physics ``Ettore Pancini'', University of Napoli and INFN-Napoli, 80126 Napoli, Italy}
\newcommand{\nagoya}{Kobayashi-Maskawa Institute for the Origin of Particles and the Universe, and Institute for Space-Earth Environmental Research, Nagoya University, Furo-cho, Chikusa-ku, Nagoya, Aichi 464-8602, Japan}
\newcommand{\laquila}{Department of Physics and Chemistry, University of L'Aquila, 67100 L'Aquila, Italy}
\newcommand{\tokyo}{Kamioka Observatory, Institute for Cosmic Ray Research, and Kavli Institute for the Physics and Mathematics of the Universe (WPI), University of Tokyo, Higashi-Mozumi, Kamioka, Hida, Gifu 506-1205, Japan}
\newcommand{\kobe}{Department of Physics, Kobe University, Kobe, Hyogo 657-8501, Japan}
\newcommand{\kit}{Institute for Astroparticle Physics \& Institute of Experimental Particle Physics, Karlsruhe Institute of Technology, 76021 Karlsruhe, Germany}
\newcommand{\tsinghua}{Department of Physics \& Center for High Energy Physics, Tsinghua University, Beijing 100084, P.R. China}
\newcommand{\ferrara}{INFN-Ferrara and Dip. di Fisica e Scienze della Terra, Universit\`a di Ferrara, 44122 Ferrara, Italy}
\newcommand{\groningen}{Nikhef and the University of Groningen, Van Swinderen Institute, 9747AG Groningen, Netherlands}
\newcommand{\westlake}{Department of Physics, School of Science, Westlake University, Hangzhou 310030, P.R. China}
\newcommand{\shenzhen}{School of Science and Engineering, The Chinese University of Hong Kong (Shenzhen), Shenzhen, Guangdong, 518172, P.R. China}
\newcommand{\coimbrapoli}{Polytechnic University of Coimbra - ISEC, 3030-199 Coimbra, Portugal}
\newcommand{\heidelberg}{Kirchhoff-Institute for Physics, Heidelberg University, 69120 Heidelberg, Germany}
\newcommand{\bucknell}{Department of Physics \& Astronomy, Bucknell University, Lewisburg, PA, USA}
\newcommand{\isct}{Department of Physics, School of Science, Institute of Science Tokyo, Meguro, Tokyo, 152-8551, Japan}
\authorrunning{XENON Collaboration}
\thankstext{addr31}{Also at \coimbrapoli}
\thankstext{xenonemail}{\texttt{xenon@lngs.infn.it}}
\thankstext{email0}{\texttt{michael.murra@columbia.edu}}
\thankstext{email1}{\texttt{guillaum@phys.columbia.edu}}
\thankstext{email2}{\texttt{masaki.yamashita@ipmu.jp}}
\institute{\hypertarget{addr3}{\columbia}\label{addr3}
\and
\hypertarget{addr28}{\groningen}\label{addr28}
\and
\hypertarget{addr23}{\tokyo}\label{addr23}
\and
\hypertarget{addr16}{\wis}\label{addr16}
\and
\hypertarget{addr17}{\zurich}\label{addr17}
\and
\hypertarget{addr18}{\paris}\label{addr18}
\and
\hypertarget{addr7}{\munster}\label{addr7}
\and
\hypertarget{addr4}{\lngs}\label{addr4}
\and
\hypertarget{addr1}{\chicago}\label{addr1}
\and
\hypertarget{addr6}{\mpik}\label{addr6}
\and
\hypertarget{addr9}{\nyuad}\label{addr9}
\and
\hypertarget{addr13}{\subatech}\label{addr13}
\and
\hypertarget{addr0}{\bologna}\label{addr0}
\and
\hypertarget{addr26}{\tsinghua}\label{addr26}
\and
\hypertarget{addr2}{\coimbra}\label{addr2}
\and
\hypertarget{addr8}{\nikhef}\label{addr8}
\and
\hypertarget{addr12}{\stockholm}\label{addr12}
\and
\hypertarget{addr30}{\shenzhen}\label{addr30}
\and
\hypertarget{addr5}{\mainz}\label{addr5}
\and
\hypertarget{addr22}{\laquila}\label{addr22}
\and
\hypertarget{addr25}{\kit}\label{addr25}
\and
\hypertarget{addr19}{\freiburg}\label{addr19}
\and
\hypertarget{addr14}{\torino}\label{addr14}
\and
\hypertarget{addr20}{\napels}\label{addr20}
\and
\hypertarget{addr15}{\ucsd}\label{addr15}
\and
\hypertarget{addr34}{\isct}\label{addr34}
\and
\hypertarget{addr21}{\nagoya}\label{addr21}
\and
\hypertarget{addr33}{\bucknell}\label{addr33}
\and
\hypertarget{addr29}{\westlake}\label{addr29}
\and
\hypertarget{addr11}{\rice}\label{addr11}
\and
\hypertarget{addr24}{\kobe}\label{addr24}
\and
\hypertarget{addr32}{\heidelberg}\label{addr32}
\and
\hypertarget{addr27}{\ferrara}\label{addr27}
}

